\documentclass[]{iopart}

\usepackage{iopams}
\usepackage{graphicx}
\usepackage{cite}

\newtheorem{definition}{Definition}[section]
\newtheorem{prop}[definition]{Proposition}

\newcommand{\hook}{\raisebox{-0.35ex}{\makebox[0.6em][r]{\scriptsize $-$}}
\hspace{-0.15em}\raisebox{0.25ex}{\makebox[0.4em][l]{\tiny $|$}}}

\begin{document}

\title[]{A deformation of Sasakian structure in the presence of torsion and supergravity solutions}

\author{Tsuyoshi Houri$^{1,2}$, Hiroshi Takeuchi$^3$ and Yukinori Yasui$^4$}

\address{$^1$%
Osaka City University Advanced Mathematical Institute (OCAMI),
3-3-138 Sugimoto, Sumiyoshi, Osaka 558-8585, Japan}
\address{$^2$%
DAMTP, Centre for Mathematical Sciences, University of Cambridge, 
Wilberforce Road, Cambridge CB3 0WA, United Kingdom}
\address{$^3$%
Department of Physics, Kyoto University, Kyoto 606-8502, Japan}
\address{$^4$%
Department of Mathematics and Physics, Graduate School of Science, Osaka City University,
3-3-138 Sugimoto, Sumiyoshi, Osaka 558-8585, Japan}

\eads{\mailto{houri@sci.osaka-cu.ac.jp}; \mailto{T.Houri@damtp.cam.ac.uk},
\mailto{takeuchi@scphys.kyoto-u.ac.jp}, \mailto{yasui@sci.osaka-cu.ac.jp}}

\begin{abstract}
A deformation of Sasakian structure 
in the presence of totally skew-symmetric torsion is discussed
on odd dimensional manifolds whose metric cones are K\"ahler with torsion.
It is shown that such a geometry inherits similar properties to those of Sasakian geometry.
As an example of them, we present an explicit expression of local metrics.
It is also demonstrated that our example of the metrics admits the existence of hidden symmetry
described by non-trivial odd-rank generalized closed conformal Killing-Yano tensors.
Furthermore, using these metrics as an {\it ansatz},
we construct exact solutions in five-dimensional minimal gauged/ungauged supergravity
and eleven-dimensional supergravity.
Finally, the global structures of the solutions are discussed.
We obtain regular metrics on compact manifolds in five dimensions,
which give natural generalizations of Sasaki-Einstein manifolds $Y^{p,q}$ and $L^{a,b,c}$.
We also briefly discuss regular metrics on non-compact manifolds in eleven dimensions.
\end{abstract}


\maketitle

%%%%%%%%%%%%%%%%%%%%%%%%%%%%%%%%%%%%%%%%%%%%%%%%%%%%%%%%%%%%%%%%%%%%%%%%%%%%%%%%%%%%%%%%%%%%%%%%%%%%
\section{Introduction}
%%%%%%%%%%%%%%%%%%%%%%%%%%%%%%%%%%%%%%%%%%%%%%%%%%%%%%%%%%%%%%%%%%%%%%%%%%%%%%%%%%%%%%%%%%%%%%%%%%%%
Sasakian geometry \cite{Sasaki:1960} has attracted intense interest
in theoretical and mathematical physics
since its applications were found in higher-dimensional supergravity theories,
string theories and M-theory.
Arguably, the most important examples are Sasaki-Einstein manifolds
which have been discussed in the context of the AdS/CFT correspondence,
especially in the physically interesting dimensions five and seven.
In five dimensions, the simplest example of the Sasaki-Einstein manifold
is the standard round five-sphere, denoted by $S^5$.
It provides a supersymmetric background $AdS_5 \times S^5$ of type IIB supergravity theory,
on which D3-brane physics is conjectured to be dual of an ${\mathcal N} = 4$
four-dimensional superconformal field theory \cite{Maldacena:1997re}.
More general five-dimensional Sasaki-Einstein manifolds $M_5$ provide
a variety of supersymmetric backgrounds $AdS_5 \times M_5$, 
which are in general dual of ${\mathcal N} = 1$ superconformal field theories.
Recently, it was proposed in \cite{ABJM:2008} that 
${\mathcal N} = 6$ three-dimensional Chern-Simons-matter theory
is related to M2-brane physics on a background $AdS_4 \times S^7/\mathbb{Z}_k$ of M-theory.
This motivates us to extend $S^7$ to general seven-dimensional Sasaki-Einstein manifolds $M_7$
and study the backgrounds $AdS_4 \times M_7$ corresponding to ${\cal N}=2$ Chern-Simons theories.
Owing to such proposals, we now have a number of concrete examples of Sasaki-Einstein manifolds.
Until recent years, the only explicit examples of Sasaki-Einstein manifolds were 
$S^5$ and $T^{1,1}$ in five dimensions and 
$M^{3,2}$, $Q^{1,1,1}$ and $V^{5,2}$ in seven dimensions.
However, thanks to Gauntlett, Martelli, Sparks and Waldram,
the infinite families of inhomogeneous Sasaki-Einstein manifolds were constructed
in five \cite{Gauntlett:2004yd,Martelli:2005} and higher \cite{Gauntlett2:2006} dimensions.
Further generalizations were constructed in various odd dimensions \cite{Cvetic:2005ft,Cvetic:2009},
in connection with vacuum rotating black hole spacetimes \cite{HashimotoEtal:2004,Chen:2006,Kubiznak:2009}.

In the familiar story of type IIB supergravity theory, 
$AdS_5 \times M_5$ backgrounds are given as supersymmetric solutions of the ten-dimensional Einstein's equation 
with the only self-dual five-form flux. 
Supersymmetry then requires $M_5$ to admit the existence of Killing spinors so that
$M_5$ is Sasaki-Einstein.
However, in general, there can be other supersymmetric solutions
which provide dual field theories still having $\mathcal{N} = 1$ supersymmetry.
Since it is expected that on these backgrounds 
one has some non-trivial fluxes which contribute to the energy-momentum tensor,
the Sasakian structure should deform.
Therefore, in order to discover such deformed backgrounds 
which, if they exist, give generalizations of the Sasaki-Einstein manifolds, 
it might be useful to think how we can deform the Sasakian structure.
Pilch and Warner \cite{Pilch:2000ej} have in fact constructed 
a non-trivial supersymmetric background $AdS_5 \times M_5$,
where $M_5$ is deformed from $S^5$ because a non-trivial three-form is present.
In \cite{CPW:2002}, a non-trivial supersymmetric background $AdS_4\times M_7$,
where $M_7$ is deformed from $S^7$, has been also constructed in M-theory.
One interesting approach in this direction is the so-called Hitchin's generalized geometry \cite{Hitchin:2003}.
By exploiting it, the notion of ``generalized Sasaki-Einstein geometry'' which
provides general supersymmetric $AdS_5$ solutions of type IIB supergravity theory 
with non-trivial fluxes
was introduced, which enables us to study the general structure of the AdS$_5$/CFT$_4$ correspondence
\cite{Gabella:2010a,Gabella:2010cy}.
Unfortunately, however, few explicit examples have been realized. 

Our aim is to deform the Sasakian structure 
by introducing a totally skew-symmetric torsion.
It is well-known that pseudo-Riemannian manifolds with totally skew-symmetric torsions
appear naturally in supergravity theories, where the torsions can be identified with
three-form or other-form fluxes occurring in the theories \cite{Strominger:1986}.
On the other hand, many kinds of torsion connections have been studied for many years
(e.g., see \cite{Agricola:2006}).
Especially in Sasakian geometry, 
the torsion connection which preserves the Sasakian structure has been studied for a long time
\cite{YAMAGucHTensor,YAMAGucHMathpura,Friedrich:2002,Friedrich2:2002}.
It is known that such a torsion is totally skew-symmetric and is written in terms of the contact one-form:
$T=\eta \wedge \textrm{d}\eta$.
The uniqueness of the torsion was proven in \cite{Friedrich:2002}.
However, since this kind of torsion connection does not deform the Sasakian structure,
we need to explore other possibilities.

In this paper, we propose one possible deformation of the Sasakian structure 
in the presence of totally skew-symmetric torsion.
The idea is the following: 
A Sasakian manifold is defined as a manifold whose metric cone in one higher dimension is K\"ahler.
Analogously, we demand that the cone one dimension up be K\"ahler with torsion.
On a K\"ahler with torsion manifold, there exists a unique torsion connection preserving the Hermitian structure,
called a {\it Bismut connection} \cite{{Bismut:1989}}, 
and the presence of the torsion deforms the K\"ahlerian structure.
Thus, Sasakian structure one dimension down is also deformed.
To our knowledge, this attempt to deform the Sasakian geometry has not been previously conceived 
and, we believe, also differs from both 
the ``generalized Sasakian geometry'' discussed in \cite{Gabella:2010a,Gabella:2010cy}
and the Sasakian geometry with torsion connection studied 
in \cite{Friedrich:2002,Friedrich2:2002}.
We thus study the general properties of the deformed Sasakian structure.

We have another motivation to study such manifolds with torsion.
It has been clarified by many authors
\cite{Cvetic:2005ft,HashimotoEtal:2004,Cvetic:2009,Kubiznak:2009,Chen:2006}
that a certain scaling limit of higher-dimensional vacuum rotating black hole solutions \cite{Myers:1986,Gibbons:2004a,Chen:2006b}
leads to toric K\"ahler metrics in even dimensions and Sasaki in odd ones.
We thus expect a similar scaling limit for charged, rotating black hole solutions
of various supergravity theories:
this leads to metrics on manifolds with torsion.
For example, it was demonstrated \cite{Houri:2012}
that in abelian heterotic supergravity, the Kerr-Sen black hole solutions \cite{Sen:1992,Cvetic2:1996,Chow2:2010} 
give rise to K\"ahler with torsion metrics.
It can be also shown that the five-dimensional gauged supergravity 
 black hole solution discovered in \cite{Chong:2005hr}
gives rise to a metric with the deformed Sasakian structure,
as we will see in section 4.

Sasakian geometry is relevant to Killing-Yano symmetry,
as exemplified by Killing-Yano (KY) tensors \cite{Yano:1952}
and conformal Killing-Yano (CKY) tensors \cite{Tachibana:1969,Kashiwada:1968,Semmelmann:2002}.
It was shown in \cite{YAMAGucHTensor, YAMAGucHMathpura} that
a Sasakian manifold of $2n+1$ dimensions has rank-$(2p+1)$
special Killing forms in the form $\eta \wedge (d\eta)^p$ ($0\leq p\leq n$).
In our case, although the Sasakian structure is deformed by torsions,
the deformed Sasakian struture admits generalized special Killing forms $\eta \wedge (d^T\eta)^p$.
Killing-Yano symmetry has also played an impartant role in the study of black hole physics.
One of the features is that general metrics admitting a rank-2 closed CKY tensor were obtained
in four \cite{Dietz:1981,Dietz:1982} 
and higher \cite{Houri:2007,Krtous:2008,Houri:2008b,Houri:2009,Frolov:2008,Yasui:2011,Santillan:2011} dimensions.
Such metrics allow remarkable properties in mathematical physics, in particular
separations of variables for the Hamilton-Jacobi, Klein-Gordon and Dirac equations.
In this paper, we present an example of the deformed Sasakian metrics explicitly.
We see that, for the example, there exists a generalized Killing-Yano symmetry
providing separability of the Hamilton-Jacobi equation for geodesics.

This paper is organized as follows.
In section 2, we begin with a brief review of torsion connections.
After we define a notion of {\it Sasaki with torsion structure}
in the presence of totally skew-symmetric torsion (see definition \ref{Def1}),
we look into general properties of the deformed Sasakian structure while
clarifying differences from the standard Sasakian structure
and introducing some new notions (see definition \ref{Def2}).
In section 3, we present an example of local metrics 
admitting the deformed Sasakian structure introduced in section 2 in all odd dimensions,
and elaborate on curvature properties with respect to the metrics and the cone metrics in one higher dimensions.
Hidden symmetry for the metrics is also discussed in this section.
In section 4,
the solutions of five-dimensional minimal (un)-gauged supergravity 
and eleven-dimensional supergravity are obtained.
In section 5,
we discuss the global structure of these solutions briefly. 
The condition to obtain regular metrics on compact manifolds are argued
in the context of five-dimensional minimal gauged supergravity solutions.
We study more on the global properties of five-dimensional solutions in the special case.
In this case, the metric has enhanced isometry and can be regarded as the generalization of $Y^{p,q}$.
Section 6 is devoted to summary and discussions.
In appendix A, we give some calculations which are relevant to the notions introduced in section 2.
In appendix B, the Riemann, Ricci and scalar curvatures 
for our example of the metrics are computed.
We get them with respect to not only the Levi-Civita connection 
but also to the connection with the torsion.
In appendix C, Calabi-Yau with torsion metrics on the cone are obtained.

%%%%%%%%%%%%%%%%%%%%%%%%%%%%%%%%%%%%%%%%%%%%%%%%%%%%%%%%%%%%%%%%%%%%%%%%%%%%%%%%%%%%%%%%%%%%%%%%%%%%
\section{Deformation of Sasakian structure}
%%%%%%%%%%%%%%%%%%%%%%%%%%%%%%%%%%%%%%%%%%%%%%%%%%%%%%%%%%%%%%%%%%%%%%%%%%%%%%%%%%%%%%%%%%%%%%%%%%%%
In the context of supergravity theories, 
it seems to be natural to introduce a totally skew-symmetric torsion
because it can be identified with three-form fields occurring in the theories \cite{Strominger:1986,Agricola:2006}.
Sasakian structure in the presence of torsion has been previously considered.
T.\ Friedrich and S.\ Ivanov \cite{Friedrich:2002,Friedrich2:2002} have used
connections with totally skew-symmetric torsion preserving Sasakian structure,
which are uniquely determined by the contact one-form $\eta$ as $T=\eta\wedge d\eta$.
On the other hand, what we expect now is that the presence of torsion no longer preserves the Sasakian structure
because of the effect of the energy-momentum tensor which changes Einstein's equation.
We thus discuss one possible deformation of the Sasakian structure 
in the presence of totally skew-symmetric torsion.

Let $(M, g)$ be a Riemannian manifold, $T$ be a 3-form on $M$
and $\{e_a\}$ be an orthonormal frame on $TM$.
A connection with totally skew-symmetric torsion $\nabla^T$
is defined by
\begin{equation}
 g(\nabla^T_X Y,Z) = g(\nabla^g_X Y,Z) + \frac{1}{2}T(X,Y,Z) ~, \label{eq00}
\end{equation}
for any vector fields $X$, $Y$ and $Z$, where $\nabla^g$ is the Levi-Civita connection of $g$.
The connection satisfies a metricity condition, $\nabla^T g=0$, 
and has the same geodesics as $\nabla^g$, 
$\nabla^T_{\dot{\gamma}}\dot{\gamma}=\nabla^g_{\dot{\gamma}}\dot{\gamma}=0$ for a geodesic $\gamma$.
The commutation relations are linked to the Lie brackets by
\begin{equation}\label{TC}
 \nabla^T_XY-\nabla^T_YX = [X,Y] + T(X,Y) ~,
\end{equation}
where
$T(X,Y,Z)=g(T(X,Y),\,Z)$.
For a $p$-form $\Psi$, the covariant derivative is calculated as
\begin{equation}
\nabla^T_X\Psi 
= \nabla^g_X\Psi-\frac{1}{2}\sum_a({X}\hook e_a\hook {T})\wedge (e_a\hook\Psi) ~, \label{eq000}
\end{equation} 
where $\hook$ represents the inner product.
Then, we have
\begin{eqnarray}
 d^T\Psi 
&= \sum_a e^a \wedge\nabla^T_{e_a}\Psi \nonumber\\
&= d\Psi - \sum_a (e_a\hook {T})\wedge(e_a\hook\Psi) ~, \\
 \delta^T\Psi 
&= -\sum_a e_a\hook \nabla^T_{e_a}\Psi \nonumber\\
&= \delta\Psi - \frac{1}{2}\sum_{a,b} 
   (e_a\hook e_b\hook {T})\wedge(e_a\hook e_b\hook\Psi) ~,
\end{eqnarray}
where $\{e^a\}$ is the dual 1-forms of $\{e_a\}$, $e_a\hook e^b=\delta^b_a$.

Suppose $(M,g,J)$ is a Hermitian manifold
equipped with a complex structure $J$
and a Hermitian metric $g$ obeying  $g(X,Y)=g(J(X), J(Y))$ for any vector field $X$ and $Y$.
Then it is known
that there exists a unique Hermitian connection $\nabla^B$
with totally skew-symmetric torsion $B$, i.e., $\nabla^B g=0$, $\nabla^B J=0$.
This connection $\nabla^B$ is known as a {\it Bismut connection}
and the corresponding totally skew-symmetric torsion $B$ is called a {\it Bismut torsion} \cite{Bismut:1989},
which is written in the form
\begin{equation}
 B(X,Y,Z)= d\Omega(J(X),J(Y),J(Z)) ~, \label{Bismut}
\end{equation}
where $\Omega$ is the fundamental 2-form $\Omega(X,Y) \equiv g(J(X), Y)$.
A Hermitian manifold $(M,g,J)$ equipped with the Bismut torsion $B$
is called a {\it K\"ahler with torsion manifold}.

A Riemannian manifold $(M, g)$ is said to be Sasakian if 
its metric cone $(C(M), \bar{g})=(M \times R_+, \bar{g}=dr^2+r^2 g)$ is K\"ahler, 
and its Sasakian structure is derived from the K\"ahler cone structure
(see, e.g., reviews \cite{Blair:1976,BoyerEtAl:2008,Sparks:2011} and references therein).
In analogy with this, we generalize the Sasakian structure
to the case when torsion is present as follows:

%%%%%%%%%%%%%%%%%%%%%%%%%%%%%%%%%%%%%%%%%%%%%%%%%%
\begin{definition} \label{Def1}
Let $(M, g)$ be a Riemannian manifold and $T$ be a 3-form on $M$.
Then, we call $(M, g, T)$ a Sasaki with torsion (ST) manifold
if its metric cone $(C(M), \bar{g})$ 
is a K\"ahler with torsion (KT) manifold
whose Bismut torsion $B$ is given by $B=r^2T$.
\end{definition}

The following propositions \ref{prop1} and \ref{prop2}
provide three equivalent characterizations of the ST structure.

%%%%%%%%%%%%%%%%%%%%%%%%%%%%%%%%%%%%%%%%%%%%%%%%%%
\begin{prop} \label{prop1}
Let $(M,g)$ be a Riemannian manifold
and $\nabla^T$ be a connection with skew-symmetric torsion $T$.
Then the following conditions are equivalent:
\begin{enumerate}
 \item[(a)] There exists a Killing vector field $\xi$ of unit length on $M$
so that the dual 1-form $\eta$ satisfies
\begin{eqnarray}
 \nabla^T_X (d^T\eta) = -2X^\flat \wedge\eta \label{COND1}
\end{eqnarray}
for any vector field $X$, where $X^\flat=g(X,-)$.
 \item[(b)] There exists a Killing vector field $\xi$ of unit length on $M$
so that the tensor field $\Phi$ of type (1,1) defined by $ \Phi(X)=\nabla^T_X \xi $ satisfies
\begin{eqnarray} 
(\nabla^T_X \Phi)(Y) = g(\xi, Y)X-g(X,Y) \xi \label{COND2}
\end{eqnarray}
for any pair of vector fields $X$ and $Y$.
 \item[(c)] There exists a Killing vector field $\xi$ of unit length on $M$ 
so that the curvature satisfies
\begin{eqnarray}
 R^T(X,Y)\xi = g(\xi,Y)X - g(\xi,X)Y + \Phi(T(X,Y)) \label{COND3}
\end{eqnarray}
for any pair of vector fields $X$ and $Y$,
where the curvature $R^T(X,Y)$ is defined by
\begin{eqnarray}
 R^T(X,Y)Z=\nabla^T_X\nabla^T_YZ-\nabla^T_Y\nabla^T_XZ-\nabla^T_{[X,Y]}Z ~. \label{curvop}
\end{eqnarray}
\end{enumerate}
\end{prop}
{\it Proof.}
$(a) \Leftrightarrow (b)$.
If $\xi$ is a Killing vector field, the dual 1-form $\eta$ satisfies
\begin{equation} \label{eta}
\nabla^T_Y \eta=\frac{1}{2}Y \hook d^T \eta
\end{equation}
for any connection with totally skew-symmetric torsion $\nabla^T$.
Since $\eta(Y)=g(\xi,Y)$, this is also written as
\begin{equation}
g(\nabla^T_Y \xi, Z)=\frac{1}{2}(d^T \eta)(Y,Z) \label{proof1}
\end{equation}
for all vector fields $Y$ and $Z$.
Thus, taking the covariant derivative of (\ref{proof1}), we have
\begin{eqnarray}
g(\nabla^T_X \nabla^T_Y \xi, Z)&=&\frac{1}{2}(\nabla^T_X d^T \eta)(Y,Z)+\frac{1}{2}(d^T\eta)(\nabla^T_X Y,Z) \nonumber\\
&=&\frac{1}{2}(\nabla^T_X d^T \eta)(Y,Z)+g(\nabla^T_{\nabla^T_X Y} \xi,Z).\label{proof2}
\end{eqnarray}
On the other hand, the covariant derivative of the equation $\nabla^T_Y \xi=\Phi(Y)$ yields
\begin{equation}
g(\nabla^T_X \nabla^T_Y \xi, Z)=g((\nabla^T_X\Phi)(Y),Z)+g(\nabla^T_{\nabla^T_X Y} \xi,Z).\label{proof3}
\end{equation}
By comparing (\ref{proof2}) and (\ref{proof3}), it follows that
\begin{equation}
(\nabla^T_X d^T \eta)(Y,Z)=2g((\nabla^T_X\Phi)(Y),Z) \,, \label{proof4}
\end{equation}
which gives the equivalence of the conditions (a) and (b).

$(b) \Rightarrow (c)$. It is noticed from (\ref{TC}) and (\ref{proof3}) that
\begin{equation}
\fl \eqalign{R^T(X,Y)\xi
&=\nabla^T_X\nabla^T_Y\xi-\nabla^T_Y\nabla^T_X\xi-\nabla^T_{[X,Y]}\xi \nonumber\\
&= (\nabla^T_X\Phi)(Y)-(\nabla^T_Y\Phi)(X)+\nabla^T_{\nabla^T_X Y} \xi
    -\nabla^T_{\nabla^T_Y X} \xi-\nabla^T_{[X,Y]}\xi \nonumber\\
&= (\nabla^T_X\Phi)(Y)-(\nabla^T_Y\Phi)(X)+\nabla^T_{T(X,Y)} \xi \,.
} \label{proof5}
\end{equation}
Since $\nabla^T_{T(X,Y)} \xi = \Phi(T(X,Y))$ by definition,
it is easy to find that the condition $(b)$ leads to the condition $(c)$.

$(b) \Leftarrow (c)$.
Using (\ref{COND3}) and (\ref{proof5}) we have
\begin{equation}
(\nabla^T_X\Phi)(Y)-(\nabla^T_Y\Phi)(X)=g(\xi,Y)X-g(\xi,X)Y \,,
\end{equation}
and also from (\ref{proof4}) we obtain
\begin{equation}
g((\nabla^T_X\Phi)(Y),Z)+g(\nabla^T_X\Phi)(Z),Y)=0 \,.
\end{equation}
Combining these two equations, we obtain (\ref{COND2}). $\square$

%%%%%%%%%%%%%%%%%%%%%%%%%%%%%%%%%%%%%%%%%%%%%%%%%%
\begin{prop} \label{prop2}
$(M,g,T)$ is an ST manifold if and only if there exists a Killing vector field $\xi$ of unit length satisfying
one of the conditions given in Prop. 2.2 and
the torsion $T$ obeys
\begin{equation}
\fl T(X,Y,Z)=T(X,\Phi(Y),\Phi(Z))+T(\Phi(X),Y,\Phi(Z))+T(\Phi(X),\Phi(Y),Z) \,. \label{torsioncond}
\end{equation}
\end{prop}
{\it Proof.}
We first derive the condition (b) in proposition \ref{prop1}
from the definition of the ST manifold,
and later show the torsion condition (\ref{torsioncond})
using the integrability of the complex structure of the metric cone.
Let $(M,g,T)$ be an ST manifold, $X$ and $Y$ be vector fields on $M$,
which can be also viewed as vector fields on the metric cone $C(M)$, 
and $\bar{\nabla}^B$ be the Bismut connection associated with $C(M)$. 
Then we have the following formulae:
\begin{equation}\label{formulae}
\eqalign{
 \bar{\nabla}^B_{\partial_r} \partial_r=0 \,, \quad
 \bar{\nabla}^B_{\partial_r} X=\bar{\nabla}^B_X \partial_r=\frac{1}{r}X \,, \\
 \bar{\nabla}^B_X Y=\nabla^T_X Y-r g(X,Y) \partial_r \,,
}
\end{equation}
where $\nabla^T$ is the connection on $M$ with totally skew-symmetric torsion $T$.
Making use of the complex structure $J$ on $C(M)$,
we define a vector field $\xi$ on $C(M)$ by 
\begin{equation}
 \xi=J(r \partial_r) \,, \label{eq2-4}
\end{equation}
whose length is given by $\bar{g}(\xi,\xi)=r^2$.
Since $\bar{\nabla}^B_X J=0$ we have
\begin{equation}
 \bar{g}(\bar{\nabla}^B_X \xi, Y)
= \bar{g}(J (\bar{\nabla}^B_X (r \partial_r)),Y)
= \bar{g}(J (X),Y) \,,
\end{equation}
which is anti-symmetric under exchange of $X$ and $Y$.
Identifying $M$ with $M \times \{1\} \subset C(M)$ leads us to the fact that
$\xi$ is a Killing vector field of unit length on $M$.
Let us define a tensor field $\Phi$ of type (1,1) by
\begin{eqnarray}\label{complex}
\Phi(X)&=&J(X)-\bar{g}(J(X), \partial_r)\partial_r \nonumber\\
&=&J(X)+r \eta(X) \partial_r \,, \label{complex}
\end{eqnarray}
where $\eta$ is the dual 1-form of the Killing vector field $\xi$, $\eta(X)=g(\xi,X)$.
Then, (\ref{COND2}) in the condition (b) follows from the covariant derivative of (\ref{complex})
and $\bar{\nabla}^B J=0$.
In fact, by virtue of the formulae (\ref{formulae}), we obtain
\begin{equation}
\eqalign{
\bar{\nabla}^B_X(\Phi(Y)) &=\nabla^T_X(\Phi(Y))-rg(X,\Phi(Y))\partial_r \\
                          &=(\nabla^T_X\Phi)(Y)+\Phi(\nabla^T_X Y)-rg(X,\Phi(Y))\partial_r \,.
}
\end{equation}
Hence the covariant derivative of $J(Y)$ is calculated as
\begin{eqnarray}\label{eq1}
\bar{\nabla}^B_X(J(Y))&=& \bar{\nabla}^B_X(\Phi(Y)-r\eta(X)\partial_r)\nonumber\\
&=&(\nabla^T_X\Phi)(Y)+\Phi(\nabla^T_XY)-\eta(\nabla^T_XY)-g(\xi,Y)X \,.
\end{eqnarray}
On the other hand we see
\begin{eqnarray}\label{eq2}
J(\bar{\nabla}^B_X Y)&=&J(\nabla^T_XY-rg(X,Y)\partial_r)\nonumber\\
&=&\Phi(\nabla^T_XY)-\eta(\nabla^T_XY)-g(X,Y)\xi \,.
\end{eqnarray}
Since $\bar{\nabla}^B J=0$, we have $\bar{\nabla}^B_X(J(Y))=J(\bar{\nabla}^B_X(Y))$
and hence equating (\ref{eq1}) and (\ref{eq2}) shows (\ref{COND2}).
Note that $\Phi(\xi)=0$ and $\Phi^2(X)=-X+\eta(X) \xi$ by the definition (\ref{complex}).
Then, (\ref{COND2}) implies that
\begin{equation}
X-g(X,\xi)\xi=(\nabla^T_X\Phi)(\xi)=-\Phi(\nabla^T_X\xi) \,.
\end{equation}
Since $\Phi^2(\nabla^T_X\xi)=-\nabla^T_X\xi$, the above equation yields
\begin{equation}
\Phi(X)=\nabla^T_X\xi \,.
\end{equation}
Thus we have obtained the condition (b) in proposition \ref{prop1}.
The torsion condition (\ref{torsioncond}) is derive
from the integrability of the complex structure $J$ on $C(M)$.
As is well known the vanishing of the Nijenhuis tensor $N_J$ of $J$
is a necessary and sufficient condition for the integrability,
so we use $N_J=0$.
From $\bar{\nabla}^B J=0$ and
\begin{equation}
[X,Y]=\bar{\nabla}^B_XY- \bar{\nabla}^B_Y X-T(X,Y) \,,\quad [X,\partial_r]=0 \,,
\end{equation}
the Nijenhuis tensor is computed as
\begin{eqnarray}
\fl \eqalign{
N_J(X,Y) &= [J(X), J(Y)]-[X,Y]-J([X,J(Y)])-J([J(X),Y]) \\
         &= -T(\Phi(X),\Phi(Y))+T(X.Y)+J(T(X,\Phi(Y))+J(T(\Phi(X),Y))
} \label{int1}
\end{eqnarray}
and
\begin{equation}\label{int2}
N_J(X,r\partial_r)=-T(\Phi(X),\xi)+J(T(X,\xi)) \,.
\end{equation}
After simple computation, the vanishing of $N_J$ ($N_J=0$) derives the condition (\ref{torsioncond}). 

Conversely, we can construct a KT structure on $C(M)$
by using the condition (a) in proposition \ref{prop1} as follows.
For the 1-form $\eta$ we introduce a 2-form $\Omega$ on $C(M)$,
\begin{equation}
\Omega=r dr \wedge \eta + \frac{r^2}{2} d^T \eta =\frac{1}{2}d^T(r^2 \eta) \,. \label{Fundform}
\end{equation}
Then the covariant derivative of $\Omega$ in radial direction always vanishes, while the derivative in direction of a vector field $X$ on $M$ yields
\begin{equation}
\fl 
\bar{\nabla}^B_X \Omega=r^2\left( X^\flat \wedge \eta+\frac{1}{2}\nabla^T_X d^T \eta \right)
+r dr \wedge \left(\nabla^T_X \eta-\frac{1}{2} X \hook d^T \eta \right)\,.
\end{equation}
The condition (a) implies the vanishing of the two brackets (for the second bracket see (\ref{eta})),
i.e., $\bar{\nabla}^B \Omega=0$.
Let us define an almost complex structure $J$ on $C(M)$ by
\begin{equation}
J(X)=\Phi(X)-r \eta(X) \partial_r \,, \quad J(r \partial_r)=\xi \,.
\end{equation}
It is easy to see that $\bar{g}(J(\bar{X}), J(\bar{Y}))=\bar{g}(\bar{X}, \bar{Y}))$
and $\Omega(\bar{X}, \bar{Y})=\bar{g}(J(\bar{X}), \bar{Y})$ for all vector fields $\bar{X}, \bar{Y}$ on $C(M)$.
Note that we have $\bar{\nabla}^BJ=0$ by $\bar{\nabla}^B\bar{g}=0$ and $\bar{\nabla}^B \Omega=0$.
For $C(M)$ to be KT, it is sufficient to show that the almost complex structure $J$ is integrable. 
It follows immediately from (\ref{int1}) and (\ref{int2}) together with the torsion condition (\ref{torsioncond}).  $\square$\\

As a consequence from propositions \ref{prop1} and \ref{prop2},
we obtain the following relations among $\xi$, $\eta$ and $\Phi$.

%%%%%%%%%%%%%%%%%%%%%%%%%%%%%%%%%%%%%%%%%%%%%%%%%%
\begin{prop} \label{prop3}
Let (M,g,T) be an ST manifold and $(\xi, \eta, \Phi)$ be a triple of its ST structure on $M$
given in proposition \ref{prop1}.
Then we have
\begin{eqnarray}
& \eta(\xi) = 1 \,, \label{AC1} \\
& \Phi(\Phi(X))=-X+\eta(X)\xi \,, \label{AC2} \\
& g(\Phi(X),\Phi(Y))=g(X,Y)-\eta(X)\eta(Y) \,, \label{AC3} \\
& \Phi(\xi)=0 \,, \quad \eta(\Phi(X))=0 \,, \label{AC4} \\
& N_{\Phi}(X,Y)+d\eta(X,Y) \xi=0 \,, \label{AC5}\\
& d^T\eta= 2\omega \,, \quad \xi \hook d\omega=0 \,,\label{AC6}
\end{eqnarray}
where $\omega$ is the fundamental 2-form defined by $\omega(X,Y)=g(\Phi(X),Y)$, and 
$N_{\Phi}$ is the Nijenhuis tensor of type-$(1,2)$ with respect to $\Phi$ defined by
\begin{equation}
\fl 
 N_{\Phi}(X,Y) = [\Phi(X), \Phi(Y)]+\Phi(\Phi([X,Y]))
                      -\Phi([X,\Phi(Y)])-\Phi([\Phi(X),Y]) \,. \label{Nijenhuis}
\end{equation}
\end{prop}

A Riemannian manifold $(M,g)$ equipped with a structure $(\xi,\eta,\Phi)$
satisfying (\ref{AC1})-(\ref{AC3})
is known as an almost contact metric manifold.
(\ref{AC4}) is derived from such a structure, especially (\ref{AC1}) and (\ref{AC2}).
An almost contact metric manifold $(M, g, \xi, \eta, \Phi)$ 
is called normal if it satisfies (\ref{AC5})
and a contact metric manifold if it satisfies $d \eta=2 \omega$, 
respectively (e.g., see \cite{BoyerEtAl:2008,Blair:1976}).
A Sasakian manifold is known as a normal contact metric manifold.
On the other hand, the ST manifold is a normal almost contact metric manifold
as the contact metric structure is deformed by the presence of torsion
as seen in (\ref{AC6}).

%%%%%%%%%%%%%%%%%%%%%%%%%%%%%%%%%%%%%%%%%%%%%%%%%%
\begin{definition} \label{Def2}
Let $(M, g)$ be a Riemannian manifold.
We call an almost contact metric structure $(g,\xi,\eta,\Phi)$ satisfying $d^T\eta= 2\omega$ and $\xi \hook d\omega=0$
together with a 3-form $T$ satisfying (\ref{torsioncond})
a T-contact metric structure, and  call $(M, g, \xi,\eta,\Phi, T)$ a T-contact metric manifold. We further call
a T-contact metric manifold a TK-contact metric manifold if $\xi$ is a Killing
vector field.
\end{definition}

An almost Cauchy-Riemann (CR) structure, which is a subbundle $E$ of $TM$ with an almost complex structure $J$,
is said to be integrable if for any sections $X, Y$ of $E$ the vector field $[J(X),Y]+[X,J(Y)]$ is a section of $E$
and the Nijenhuis tensor of $J$ vanishes.
The subbundle $\mathcal{D}=$ker $\eta \subset TM$ has an almost complex structure defined by
$J_{\mathcal{D}}=\Phi |_{\mathcal{D}}$. Hence, $\mathcal{D}$ together with the endomorphism 
$J_{\mathcal{D}}$ provides $M$ with an almost CR structure of codimension one. 
The normality condition yields that the almost CR structure $(\mathcal{D},J_{\mathcal{D}})$ is integrable.

%%%%%%%%%%%%%%%%%%%%%%%%%%%%%%%%%%%%%%%%%%%%%%%%%%
\begin{prop} \label{prop4}
An ST manifold is a normal T-contact metric manifold whose torsion 
$T_{\mathcal{D}}=T |_{\mathcal{D}}$
is given by a Bismut torsion
\begin{equation}\label{Bis}
T_{\mathcal{D}}(X, Y, Z)=d \omega(J_{\mathcal{D}}(X), J_{\mathcal{D}}(Y),J_{\mathcal{D}}(Z)) 
\end{equation}
for all $X,Y,Z \in \mathcal{D}$. 
\end{prop}
{\it Proof.}
Let $(M,g,T)$ be an ST manifold. 
Since we find from proposition \ref{prop3} that $M$ is a normal $T$-contact metric manifold,
we have $N^{(i)}=0$ $(i=1,2)$ where $N^{(i)}$ are defined by (\ref{N1}) and (\ref{N2}).
Then, (\ref{dg}) reduces to
\begin{equation} \label{LKT0}
\fl \eqalign{
 2 g((\nabla^T_X \Phi)(Y), Z) =& -d \omega(X, \Phi (Y), \Phi (Z))+d \omega(X,Y,Z)+M(X,Y,Z) \\
                             & +d^T \eta(X, \Phi (Z)) \eta(Y)-d^T \eta(X, \Phi (Y)) \eta(Z) \,,
}
\end{equation}
Using (\ref{COND2}) and
\begin{eqnarray}
&d^T \eta(X, \Phi (Z)) \eta(Y)-d^T \eta(X, \Phi (Y)) \eta(Z) \nonumber\\
&=2 \omega(X, \Phi (Z)) \eta(Y)-2 \omega(X, \Phi (Y)) \eta(Z) \nonumber\\
&=2 g(X, Z) g(\xi,Y)-2 g(X, Y) g(\xi,Z) \,,
\end{eqnarray}
we obtain
\begin{equation} \label{LKT}
d \omega(X, \Phi Y, \Phi Z)-d \omega(X,Y,Z)-M(X,Y,Z)=0 \,.
\end{equation}
From (\ref{AC6}) and (\ref{M}), it holds trivially
if we take $X=\xi$, $Y=\xi$ or $Z=\xi$.
Otherwise, (\ref{LKT}) is equivalent to (\ref{Bis}) for $X$, $Y$ and $Z \in \mathcal{D}$.

Conversely, the normality condition $N^{(1)}=0$ leads to $\mathcal{L}_\xi \Phi=0$ (see \cite{BoyerEtAl:2008}), 
so that $\xi$ is a Killing vector field (see (\ref{Kill})).
Following the same calculation as (\ref{LKT0})--(\ref{LKT}) inversely,
we obtain the condition (b) in proposition \ref{prop1}. $\Box$\\
%%%%%%%%%%%%%%%%%%%%%%%%%%%%%%%%%

Since an almost contact metric structure is normal 
if and only if the almost CR structure is integrable and $\mathcal{L}_\xi \Phi=0$ (see \cite{BoyerEtAl:2008}),
we are able to restate proposition \ref{prop4} in the following proposition.

%%%%%%%%%%%%%%%%%%%%%%%%%%%%%%%%%%%%%%%%%%%%%%%%%%
\begin{prop} \label{prop5}
An ST manifold is a TK-contact metric manifold whose almost CR structure is integrable and  torsion 
$T_{\mathcal{D}}=T |_{\mathcal{D}}$
is given by a Bismut torsion.
\end{prop}
%%%%%%%%%%%%%%%%%%%%%%%%%%%%%%%%%%%%%%%%%%%%%%%%%%

Let us close this section by mentioning about some other properties of the ST manifolds.
A $p$-form $\phi$ is called a {\it special Killing $p$-form with torsion}
if it satisfies for any vector field $X$
\begin{eqnarray} \label{Special}
 \nabla^T_X \phi = \frac{1}{p+1} X \hook d^T \phi \,, \quad
 \nabla^T_X (d^T \phi) = k\,X \wedge \phi 
\end{eqnarray}
with a constant $k$.
For $\phi=\eta$, the first equation implies 
that its dual vector field $\xi$ is a Killing vector field.
Hence, the 1-form $\eta$ in proposition \ref{prop1} is a special Killing 1-form with torsion.
Furthermore, it can be shown \cite{Houri:2012} that the $(2\ell+1)$-forms
\begin{eqnarray}
\eta^{(\ell)}=\eta \wedge (d^T \eta)^{\ell} \label{etaell}
\end{eqnarray}
for $\ell=0,\cdots,n$, are also special Killing forms with torsion.
For a special Killing $p$-form with torsion $\phi$ on $M$,
\begin{equation}
\hat{\phi}=r^p dr \wedge \phi+\frac{r^{p+1}}{p+1} d^T \phi
\end{equation}
is a parallel $(p+1)$-form on $C(M)$, i.e., $\bar{\nabla}^B \hat{\phi}=0$ (see \cite{Semmelmann:2002}).
In particular, for $p=1$, the 1-form $\eta$ on an ST manifold provides
a parallel 2-form $\Omega$ on $C(M)$,
which is precisely a fundamental 2-form on $C(M)$ (cf.\ (\ref{Fundform})).

It is known that the Ricci tensor of a Sasakian manifold of dimension $2n + 1$ is given by
$\textrm{Ric}(X,\xi) = 2n\,\eta(X)$.
In the ST manifold case, the Ricci curvature follows from (\ref{COND3}) that
\begin{eqnarray}
 \textrm{Ric}^T(X,\xi)
&= -\sum_a g(R^T(X,e_a)\xi,e_a) \nonumber\\
&= 2n \,\eta(X) - \sum_a T(X,e_a,\Phi(e_a))  ~.
\end{eqnarray}

%%%%%%%%%%%%%%%%%%%%%%%%%%%%%%%%%%%%%%%%%%%%%%%%%%%%%%%%%%%%%%%%%%%%%%%%%%%%%%%%%%%%%%%%%%%%%%%%%%%%
\section{Sasaki with torsion metrics}
%%%%%%%%%%%%%%%%%%%%%%%%%%%%%%%%%%%%%%%%%%%%%%%%%%%%%%%%%%%%%%%%%%%%%%%%%%%%%%%%%%%%%%%%%%%%%%%%%%%%
It would be useful to give some examples of the ST manifolds explicitly,
as many examples of the Sasakian manifolds have been used for tests of AdS/CFT correspondence.
In what follows we shall discuss a concrete example of the ST metric
which possesses the general properties of the ST structure we have already seen in section 2.
The metric contains some unknown functions of single variable,
which are determined by equations of motion of supergravity theories in section 4
and further restricted by regularity conditions in section 5.
We proceed the calculation in this section while keeping the single variable functions unknown.
In section 3.1, we give a physical motivation to consider our example especially in supergravity theories.
In section 3.2, we confirm that the cone metric of our example is K\"ahler with torsion
and then give the relation between the torsion of the ST and the Bismut torsion of the cone.
In our case it is also found that the metric possesses Killing-Yano symmetry
which is described by a generalized closed conformal Killing-Yano 3-form.
To our knowledge, it is the first example of the metric admitting such a 3-form.
Therefore, we investigate in section 3.3 some properties of the ST metric
from the view point of Killing-Yano symmetry.

%%%%%%%%%%%%%%%%%%%%%%%%%%%%%%%%%%%%%%%%%%%%%%%%%%
\subsection{Local metrics in all odd dimensions}
%%%%%%%%%%%%%%%%%%%%%%%%%%%%%%%%%%%%%%%%%%%%%%%%%%
It has been realized \cite{HashimotoEtal:2004,Cvetic:2005ft,Cvetic:2009,Kubiznak:2009}
that the well-known examples of the toric Sasakian manifolds
such as $Y^{p,q}$ and $L^{a,b,c}$, originally constructed by \cite{Gauntlett:2004yd,Gauntlett2:2006,Martelli:2005},
can be obtained by taking the BPS limit of the Euclidean vacuum rotating black hole solutions
in five and higher dimensions.
The general Sasakian metric in $2n+1$ dimensions is locally written
as an $S^1$-bundle over $2n$-dimensional K\"ahler space $(B,g_B)$,
\begin{eqnarray}
g = g_{B} + 4(d\psi_0 + {\cal A})^2 \,.
\end{eqnarray}
Since a lot of charged black hole solutions of the equations of motion in supergarvity theories
have been discovered,
it naturally motivates us to ask what happens
when we start with charged black holes in supergravity theories.

We shall explicitly present an example 
of local metrics admitting the deformed Sasakian structure introduced in section 2,
which we call Sasaki with torsion (ST) metrics.
The ST metric in $2n+1$ dimensions we found is given
in local coordinates ($x^a$)=($x_\mu,\psi_k$) 
where $\mu=1,\cdots,n$ and $k=0,\cdots,n$, by
\begin{equation}
\fl g = \sum_{\mu=1}^n \frac{dx_\mu^2}{Q_\mu}
              +\sum_{\mu=1}^n Q_\mu \left( \sum_{k=1}^n \sigma^{(k-1)}_\mu d\psi_k \right)^2
             +4\left(\sum_{k=0}^n \sigma^{(k)} d\psi_k+A \right)^2 \,, \label{met}
\end{equation}
where
\begin{equation}
\fl A = \sum_{\mu=1}^n \frac{N_\mu}{U_\mu}\sum_{k=1}^n \sigma_\mu^{(k-1)} d\psi_k \,, \quad
    Q_\mu = \frac{X_\mu}{U_\mu} \,, \quad
    U_\mu = \prod_{\nu=1, \nu\neq\mu}^n(x_\mu-x_\nu) \label{funcs}
\end{equation}
and $\sigma_\mu^{(k)}$ and $\sigma^{(k)}$ are the $k$-th elementary symmetric polynomials in $x_\mu$
generated by
\begin{equation}
\fl \prod_{\nu=1, \nu\neq\mu}^n(\lambda+x_\nu) = \sum_{k=0}^{n-1}\sigma_\mu^{(k)}\lambda^{n-k-1} \,, \quad
 \prod_{\nu=1}^n (\lambda + x_\nu) = \sum_{k=0}^n \sigma^{(k)}\lambda^{n-k} \,.
\end{equation}
The metric contains $2n$ unknown functions $X_\mu(x_\mu)$ and $N_\mu(x_\mu)$
depending only on single variable $x_\mu$.
Although the unknown functions are determined 
by the equations of motion of various supergravity theories and the regularity of the metric
as we will see in section 4 and 5,
we proceed the calculation keeping them arbitrary in this section.
It is known \cite{Cvetic:2005ft,HashimotoEtal:2004,Cvetic:2009,Kubiznak:2009,Chen:2006}
that the metric (\ref{met}) with $A=0$ is obtained 
as an ``off-shell'' metric of the BPS limit of the odd-dimensional Kerr-NUT-(A)dS metric
and leads to the toric Sasaki-Einstein metrics $Y^{p,q}$ and $L^{a,b,c}$ discovered by
\cite{Gauntlett:2004yd,Gauntlett2:2006,Martelli:2005}.
According to proposition \ref{prop4} (or \ref{prop5}),
it implies that the metric $g_B$ on $2n$-dimensional base space $(B,g_{B})$ is locally KT.
the present metric is known as an orthotoric K\"ahler metric
established in \cite{Apostolov:2004,Apostolov:2006}.

For later calculation, it is convenient to introduce an orthonormal frame 
$\{e^a\}=\{e^\mu,e^{\hat{\mu}}=e^{n+\mu},e^0=e^{2n+1}\}$.
We choose an orthonormal frame for the metric (\ref{met}) as
\begin{equation}
\eqalign{
 e^\mu = \frac{dx_\mu}{\sqrt{Q_\mu}} \,, \quad
 e^{\hat{\mu}} = \sqrt{Q_\mu}\sum_{k=1}^n \sigma_\mu^{(k-1)} d\psi_k \,, \\
 e^0 = 2\Big(\sum_{k=0}^n \sigma^{(k)} d\psi_k+A\Big) \,.
}\label{ortho}
\end{equation}
From the first structure equation
\begin{eqnarray}
 de^a + \sum_b \omega^a{}_b\wedge e^b = 0
\end{eqnarray}
and $\omega_{ab}=-\omega_{ba}$,
we compute the connection 1-forms $\omega^a{}_b$ as follows:
\begin{equation}
\eqalign{
 \omega^\mu{}_\nu 
=& -\frac{\sqrt{Q_\nu}}{2(x_\mu-x_\nu)}\,e^\mu
                          -\frac{\sqrt{Q_\mu}}{2(x_\mu-x_\nu)}\,e^\nu \,,\quad (\mu\neq\nu) \label{con1} \\
 \omega^\mu{}_{\hat{\mu}} 
=& -\partial_\mu\sqrt{Q_\mu}\,e^{\hat{\mu}}
                               +\sum_{\nu\neq\mu}\frac{\sqrt{Q_\nu}}{2(x_\mu-x_\nu)}e^{\hat{\nu}}
                                -(1+\partial_\mu H)\,e^0 \,, \\
 \omega^\mu{}_{\hat{\nu}}
=& \frac{\sqrt{Q_\nu}}{2(x_\mu-x_\nu)}\,e^{\hat{\mu}}
                          -\frac{\sqrt{Q_\mu}}{2(x_\mu-x_\nu)}\,e^{\hat{\nu}} \,, \quad (\mu\neq\nu) \\
 \omega^{\hat{\mu}}{}_{\hat{\nu}}
=& -\frac{\sqrt{Q_\nu}}{2(x_\mu-x_\nu)}\,e^\mu
                          -\frac{\sqrt{Q_\mu}}{2(x_\mu-x_\nu)}\,e^\nu \,, \quad (\mu\neq\nu) \\
 \omega^\mu{}_0 
=& -(1+\partial_\mu H)\,e^{\hat{\mu}} \,, \\
 \omega^{\hat{\mu}}{}_0 
=& (1+\partial_\mu H)\,e^\mu \,,
} \label{con6}
\end{equation}
where $H$ is defined by
\begin{equation}
H=\sum_{\mu=1}^n \frac{N_\mu}{U_\mu} ~. \label{functionH}
\end{equation}

Firstly we shall see the conditions in proposition \ref{prop3}.
We introduce a 1-form $\eta$, vector field $\xi$ and endmorphism $\Phi$ as
\begin{equation}
 \eqalign{
 \eta= e^0 \,, \quad \xi= e_0 \,, \\
 \Phi(e_\mu)=e_{\hat{\mu}} \,, \quad \Phi(e_{\hat{\mu}})=-e_\mu \,,\quad  \Phi(e_0)=0 \,. }
\end{equation}
For the triple $(\xi,\eta,\Phi)$ together with the metric $g$,
the conditions (\ref{AC1})--(\ref{AC3}) in proposition \ref{prop3} hold clearly,
so that $(g,\xi,\eta,\Phi)$ is an almost contact metric structure.
Using $e_c\hook\nabla_{e_a}e^b=-\omega^b{}_c(e_a)$ 
we compute the covariant derivatives with respect to the Levi-Civita connection $\nabla^g$
as (\ref{CovDer1-13}) in appendix B, 
and its commutation relations $[e_a,e_b]$ are obtained.
From the obtained commutation relations
we are able to confirm the condition (\ref{AC5}), 
which means that the almost contact metric structure is normal.
However, $\eta$ is not in general a contact 1-form because we have
\begin{eqnarray}
 d\eta = 2\sum_{\mu=1}^n (1+\partial_\mu H)\,e^\mu\wedge e^{\hat{\mu}} ~,
\end{eqnarray}
and hence there is a possibility that $\eta \wedge (d\eta)^n=0$ at some points.
If $H$ is constant, we have $d\eta = 2\omega$, where $\omega$ is the fundamental form,
so that $\eta$ is a contact 1-form.
It is also found that the present metric is a quasi-Sasakian metric \cite{Blair:1967}, 
whose fundamental form satisfies $d\omega = 0$.
In fact, we have
\begin{eqnarray}
 \omega = \sum_{\mu=1}^n e^\mu\wedge e^{\hat{\mu}} 
        = d\Bigg[\sum_{k=0}^n\sigma^{(k)}d\psi_k\Bigg] \,. \label{fun2}
\end{eqnarray}

Next, let us see the conditions in proposition \ref{prop2}.
We introduce the torsion $T$ and compute the covariant derivatives
with respect to the torsion connection $\nabla^T$.
Since the torsion $T$ satisfying (\ref{AC6}) is given by
\begin{eqnarray}
 T = 2 \sum_{\mu=1}^n \partial_\mu H \,e^\mu \wedge e^{\hat{\mu}} \wedge e^0 ~, \label{torsion}
\end{eqnarray}
we can check that (\ref{torsioncond}) holds.
We emphasize again that the torsion (\ref{torsion})
differs from the torsion preserving the Sasakian structure, $\eta\wedge d\eta$,
discussed in \cite{Friedrich:2002}.
Namely, $\nabla^T\xi\neq 0$, $\nabla^T\eta\neq 0$ and $\nabla^T\Phi\neq 0$.
The covariant derivatives with respect to $\nabla^T$ are calculated 
as (\ref{CovDer2-13}) in appendix B.
Using these expressions, we find that 
\begin{eqnarray}
 \nabla^T_X\xi = \Phi(X) ~.
\end{eqnarray}
It is also shown that for any vector field $X$,
\begin{eqnarray}
 \nabla^T_X \eta = \frac{1}{2} X \hook d^T \eta ~,~~~
 \nabla^T_X (d^T \eta) = -2 X \wedge \eta ~,
\end{eqnarray}
which proves (\ref{Special}) with $k=-2$ so that $\eta$ is a special Killing 1-form with torsion.

%%%%%%%%%%%%%%%%%%%%%%%%%%%%%%%%%%%%%%%%%%%%%%%%%%
\subsection{The cone metric}
%%%%%%%%%%%%%%%%%%%%%%%%%%%%%%%%%%%%%%%%%%%%%%%%%%
Going back to the definition \ref{Def1},
we confirm that the Riemannian cone metric of our example is K\"ahler with torsion.
For later calculation, we introduce an orthonormal frame $\{\bar{e}^\alpha\}$
$(\alpha=r,1,\cdots,n)$
\begin{eqnarray}
\bar{e}^r = dr \,, \quad \bar{e}^a=r e^a \label{ortho_cone}
\end{eqnarray}
with respect to the cone metric
\begin{equation}
\bar{g}=dr^2+r^2g \label{met_cone}
\end{equation} 
where $g$ is given by (\ref{met}).

The connection 1-forms $\bar{\omega}^\alpha{}_\beta$ with respect to $\bar{g}$
are calculated as
\begin{eqnarray}
\bar{\omega}^r{}_a=-\frac{1}{r} \bar{e}^a ~,~~~ 
\bar{\omega}^a{}_b=\omega^a{}_b ~, \label{coneomega}
\end{eqnarray}
where $\omega^a{}_b$ is given by (\ref{con1})--(\ref{con6}),
and the commutation relations $[\bar{e}_\alpha,\bar{e}_\beta]$
are calculated in the similar manner to previous section.
We introduce an almost complex structure $J$ by
\begin{eqnarray}
 J(\bar{e}_r)=\bar{e}_0 ~,~~~ J(\bar{e}_0)=-\bar{e}_r ~,~~~
 J(\bar{e}_\mu)=\bar{e}_{\hat{\mu}} ~,~~~ J(\bar{e}_{\hat{\mu}})=-\bar{e}_\mu ~.
\end{eqnarray}
Then it is directly checked that for the almost complex structure $J$, 
the Nijenhuis tensor vanishes so that $J$ is integrable,
and the cone metric $\bar{g}$ is Hermitian,
\begin{eqnarray}
 \bar{g}(X,Y)=\bar{g}(J(X), J(Y)) ~.
\end{eqnarray}
The fundamental form $\Omega(X,Y)=\bar{g}(J(X),Y)$ can be written as
\begin{eqnarray}
\Omega = \bar{e}^r \wedge \bar{e}^0+\sum_{\mu=1}^n \bar{e}^\mu \wedge \bar{e}^{\hat{\mu}}
       = \frac{1}{2} d^T (r^2 e^0) ~. 
\end{eqnarray}
Since ($M,g,J$) is a Hermitian manifold, there exists the Bismut connection, 
a unique Hermitian connection $\bar{\nabla}^B$
with totally skew-symmetric torsion $B$.
From (\ref{Bismut}) the Bismut torsion is explicitly obtained as
\begin{eqnarray}
 B = \sum_{\mu=1}^n \frac{2}{r} \,\partial_\mu H
          \,\bar{e}^\mu \wedge \bar{e}^{\hat{\mu}} \wedge \bar{e}^0  = r^2 T ~, \label{B_cone}
\end{eqnarray}
where $T$ is given by (\ref{torsion}).
We finally note that the Killing vector fields $ \partial/\partial \psi_k~~(k=0, 1, \cdots, n)$
preserve the KT structure on the cone,
\begin{equation}\label{KILL}
\mathcal{L}_{\partial_k} \Omega=0,~~\mathcal{L}_{\partial_k} B=0.~~ 
\end{equation}

%%%%%%%%%%%%%%%%%%%%%%%%%%%%%%%%%%%%%%%%%%%%%%%%%%
\subsection{Hidden symmetry}
%%%%%%%%%%%%%%%%%%%%%%%%%%%%%%%%%%%%%%%%%%%%%%%%%%
It is known that a generalized Killing-Yano symmetry in the presence of totally skew-symmetric torsion, which were introduced by \cite{Kubiznak:2009qi},
appears for the black hole solutions of the five-dimensional minimal gauged supergravity \cite{Kubiznak:2009qi}
and of the abelian heterotic supergravity in four \cite{Sen:1992}, five \cite{Cvetic2:1996} and higher dimensions \cite{Chow2:2010}.
Moreover, it has been realized lately
that the generalized Killing-Yano symmetry is related to 
K\"ahler manifolds established  by \cite{Apostolov:2006,Apostolov:2004} and toric Sasakian manifolds
which are obtained as the BPS limit of Euclideanized higher-dimensional black hole spacetimes
\cite{Houri:2012,Papadopoulos:2011a,Papadopoulos:2011b,Visinescu:2012}.
As we have seen in previous sections, the ST metric (\ref{met}) 
can be regarded as a natural generalization of Sasakian metrics in the presence of torsion.
Since the ordinary Sasakian metric obtained from vacuum black holes admits the generalized Killing-Yano symmetry,
it is natural to expect that the ST metric (\ref{met}) also admits the generalized Killing-Yano symmetry.

A {\it generalized conformal Killing-Yano (GCKY) tensor $k$} was introduced by \cite{Kubiznak:2009qi}
as a $p$-form satisfying for any vector field $X$ and a totally skew-symmetric torsion $T$,
\begin{eqnarray}
\nabla^T_X k
= \frac{1}{p+1}X\hook d^Tk 
  - \frac{1}{D-p+1}X^\flat\wedge\delta^T k ~, \label{1-5}
\end{eqnarray}
where $X^\flat$ is the dual 1-form of $X$.
In particular, a GCKY tensor $f$ obeying $\delta^Tf=0$ is called a {\it generalized Killing-Yano (GKY) tensor},
and a GCKY tensor $h$ obeying $d^Th=0$ is called a {\it generalized closed conformal Killing-Yano (GCCKY) tensor}.
From general properties \cite{Houri:2010fr,HouriEtal:2010,KubiznakEtal:2010}, any GKY tensors $f$ of rank-$p$ always provide
rank-2 Killing tensors $K$ obeying $\nabla_{(a}K_{bc)}=0$, by
\begin{eqnarray}
 K_{ab} = f_{ac_1\cdots c_{p-1}}f_b{}^{c_1\cdots c_{p-1}} \,. \label{KYtoKT}
\end{eqnarray}
When a Hamilton-Jacobi equation for geodesics
can be solved by separation of variables,
the separation constants $\kappa^{(i)}$ are given
as the eigenvalues of rank-2 Killing tensors $K^{(i)}$, $\kappa^{(i)}=K^{(i)}{}^{ab}p_ap_b$.
Hence, the separability of Hamilton-Jacobi equations for geodesics provides rank-2 Killing tensors.
On the other hand, not all the rank-2 Killing tensors can be decomposed
into the square of Killing-Yano tensors as (\ref{KYtoKT}).
Nevertheless, it is easy to demonstrate that 
for the metric (\ref{met}) the Hamilton-Jacobi equation for geodesics separates,
and we obtain rank-2 Killing tensors.
Therefore, it is an interesting problem to investigate
whether the Killing tensors are given by GKY tensors or not.

To explore the GKY tensors for the metric (\ref{met}), we have to determine a torsion connection first. 
The natural torsion is the 3-form $T$ related to the ST structure, given by (\ref{torsion}).
Since the first equation in (\ref{Special}) is same as the GKY equation,
a special Killing $p$-form with torsion is alternatively said 
to be a {\it rank-$p$ special GKY tensor}.
As was already seen in (\ref{etaell}), $\eta^{(\ell)}\equiv \eta \wedge (d^T \eta)^{\ell}$
for $\ell=0,\cdots,n$ are rank-$(2\ell+1)$ special GKY tensors
with respect to torsion $T$.
Thus we have $n+1$ GKY tensors.
However, these GKY tensors $\eta^{(\ell)}$ do not give rise 
to non-trivial rank-2 Killing tensors.
In fact, every GKY tensor generates the only metric essentially.

Introducing another torsion, we find other GKY tensors $f^{(j)}$ for the metric (\ref{met}),
which are not special GKY. 
We introduce a 2-form $\hat{h}$ and 3-form $G$ as
\begin{eqnarray}
 \hat{h} =& \sum_{\mu=1} \sqrt{x_\mu} \,e^\mu\wedge e^{\hat{\mu}} ~, \\
 G =& \sum_{\mu\neq\nu}\frac{1}{\sqrt{x_\mu}+\sqrt{x_\nu}}\sqrt{\frac{Q_\nu}{x_\nu}}
      \,e^\mu\wedge e^{\hat{\mu}}\wedge e^{\hat{\nu}} \nonumber\\
      &+\sum_{\mu=1}^n 2(1+\partial_\mu H)
      \,e^\mu\wedge e^{\hat{\mu}}\wedge e^0 ~. \label{GCKYtorsion}
\end{eqnarray}
Then it is demonstrated that for the metric (\ref{met}), the $(2j+1)$-forms
\begin{eqnarray}
 h^{(\ell)} \equiv e^0 \wedge (\hat{h})^j
         = e^0 \wedge \hat{h} \wedge \cdots \wedge \hat{h} \label{defH}
\end{eqnarray}
for $j=1,\cdots,n$, are rank-$(2j+1)$ GCCKY tensors with respect to torsion $G$, 
obeying for any vector field $X$
\begin{eqnarray}
 \nabla^G_X h^{(j)} = -\frac{1}{D-2j}X^\flat \wedge \delta^G h^{(j)} ~.
\end{eqnarray}
From general properties of GCKY tensors (e.g., see \cite{Houri:2010fr}),
GCCKY tensors $h^{(j)}$ generate GKY tensors $f^{(j)}$ by $f^{(j)}=*h^{(j)}$.
These GKY tensors $f^{(j)}$ generate rank-2 Killing tensors $K^{(j)}$ by
$K^{(j)}_{ab} = [\ell\,!^2(n-2j-1)\,!]^{-1}\, f^{(j)}{}_{ac_1\cdots c_{D-2\ell-2}}
f^{(j)}{}_b{}^{c_1\cdots c_{D-2\ell-2}}$,
which are explicitly written as
\begin{eqnarray}
 K^{(j)} = \sum_{\mu=1}^n \sigma_\mu^{(j)}\,(e^\mu\otimes e^\mu+e^{\hat{\mu}}\otimes e^{\hat{\mu}}) ~.
\end{eqnarray}

%%%%%%%%%%%%%%%%%%%%%%%%%%%%%%%%%%%%%%%%%%%%%%%%%%%%%%%%%%%%%%%%%%%%%%%%%%%%%%%%%%%%%%%%%%%%%%%%%%%%
\section{Supergravity solutions}
%%%%%%%%%%%%%%%%%%%%%%%%%%%%%%%%%%%%%%%%%%%%%%%%%%%%%%%%%%%%%%%%%%%%%%%%%%%%%%%%%%%%%%%%%%%%%%%%%%%%
In the context of supergravity theories, 
it seems to be natural to introduce a totally skew-symmetric torsion
because it can be identified with three-form fields
occurring in the theories \cite{Strominger:1986,Agricola:2006}.
In this section we investigate Euclidean solutions of two particular supergravity theories,
the five-dimensional gauged minimal supergravity in section 4.1
and the eleven-dimensional supergravity in section 4.2.
As mentioned before, there is a correspondence between Kerr-dS black holes
and toric Sasakian manifolds,
which can be seen through a Wick rotation and a certain scaling limit.
Analogously, it is expected to obtain an Euclidean solution
corresponding to the charged Kerr-dS black hole solution \cite{Chong:2005hr}
in the five-dimensional gauged minimal supergravity.
By making use of the canonical form (\ref{met}) for the ST metric in section 3,
we attempt to solve equations of motion of the theory.
Similarly to five dimensions, we also explore an Euclidean solution
of the eleven-dimensional supergravity under the same {\it ansatz},
because it is suggested that there are a lot of similarities between
the five and eleven-dimensional supergravities.
Since any charged, rotating black hole solution is not known, if exists,
the Euclidean solution might give us a clue for finding new black hole solution
in eleven-dimensional supergravity.

%%%%%%%%%%%%%%%%%%%%%%%%%%%%%%%%%%%%%%%%%%%%%%%%%%
\subsection{Five-dimensional minimal gauged supergravity}
%%%%%%%%%%%%%%%%%%%%%%%%%%%%%%%%%%%%%%%%%%%%%%%%%%
The five-dimensional minimal gauged supergravity is given
by the (Lorentzian) action
\begin{equation}
S_5 = \int *({\cal R}-\Lambda)-\frac{1}{2} F_{(2)}\wedge *F_{(2)}
        +\frac{1}{3\sqrt{3}} F_{(2)}\wedge F_{(2)}\wedge A_{(1)} ~,
\end{equation}
where $F_{(2)}=dA_{(1)}$ is a 2-form field strength of a Maxwell field $A_{(1)}$,
${\cal R}$ is the Ricci curvature of a gravitational field $g_5$
and $\Lambda$ is the cosmological constant.
The equations of motion are given by
\begin{eqnarray}
& R_{ab} = -4 g_{ab} + \frac{1}{2}\Big(F_{(2)ac}F_{(2)b}{}^c
           -\frac{1}{6}g_{ab}F_{(2)cd}F_{(2)}^{cd}\Big) \,, \label{5dEin}\\
& d* F_{(2)}-\frac{1}{\sqrt{3}} F_{(2)}\wedge F_{(2)} = 0 \,, \label{5dCS}
\end{eqnarray}
where the cosmological constant has been normalized as $\Lambda = -12$.

It should be noted here that for Euclidean solutions, 
we must consider the Euclidean action which is obtained by the Wick rotation.
Since it corresponds to change the sign of the whole right-hand side of (\ref{5dEin}),
the cosmological constant can be interpreted as positive.
The Wick rotation we take transforms the only fiber direction from spacelike into timelike,
so as to satisfy the original Einstein equation (\ref{5dEin}),
and does not break the reality of the matter flux.
Therefore, we investigate the Einstein equation for Euclidean signature,
\begin{eqnarray}
 R_{ab} = 4 g_{ab} - \frac{1}{2}\Big(F_{(2)ac}F_{(2)b}{}^c
           -\frac{1}{6}g_{ab}F_{(2)cd}F_{(2)}^{cd}\Big) \,. \label{Euc5dEin}
\end{eqnarray}

As for the gauge potential $A_{(1)}$ and the functions $N_\mu$,
we assume the following form so as to solve the Maxwell Chern-Simons equation (\ref{5dCS}),
\begin{eqnarray}
A_{(1)} = &c_F \sum_{\mu=1}^2 \frac{q_\mu}{U_\mu}\sum_{k=1}^2 \sigma_\mu^{(k-1)} d\psi_k ~,\label{vec} \\
N_\mu = &a_1 x_\mu + q_\mu ~,
\end{eqnarray}
with constant parameters $c_F$, $a_1$ and $q_\mu$.
Since $a_1$ is a gauge parameter, we set $a_1=0$.
In the form, the field strength is given by
\begin{equation}
 F_{(2)} = c_F \left(\partial_1H \,e^1\wedge e^{\hat{1}}
                         +\partial_2H \,e^2\wedge e^{\hat{2}} \right) \,, \label{Mxl}
\end{equation}
where $H$ is, as before, given by (\ref{functionH}).
This immediately shows that $\partial_1H = -\partial_2H$ and hence
\begin{eqnarray}
 * F_{(2)} = - F_{(2)} \wedge \eta \,, \qquad
 F_{(2)} \wedge \omega  =0 \,,
\end{eqnarray}
where $\eta$ is the contact one-form and $\omega$ is the fundamental 2-form given by (\ref{fun2}).
Thus, the Maxwell equation (\ref{5dCS}) can be solved easily as
\begin{equation}
 \textrm{d}*F_{(2)}
= - F_{(2)} \wedge \textrm{d}\eta
= -\frac{2}{c_F}\,F_{(2)}\wedge F_{(2)} \,,
\end{equation}
where the constant $c_F$ is determined as $c_F=-2\sqrt{3}$.
The Einstein equation (\ref{Euc5dEin}) requires that $X_\mu(x_\mu)$ takes the form
\begin{eqnarray}
 X_\mu=& -4 x_\mu^3+\sum_{i=1}^2 c_i x_\mu^i + b_\mu- 8 q_\mu x_\mu ~, \label{eq4-10a}
\end{eqnarray}
where $c_i$, $b_\mu$ and $q_\mu$ are constants.

Finally let us comment a solution of the five-dimensional ungauged minimal supergravity.
We obtain an ungauged minimal supergravity solution in the similar way.
The solution is provided when (\ref{met}) and  (\ref{vec}) take the form
\begin{eqnarray}
 X_\mu= \sum_{i=1}^{2} c_i x_\mu^i + b_\mu ~,~~~ 
 N_\mu = - x_\mu ^2 + a_1 x_\mu + q_\mu ~.
\end{eqnarray}
The solutions can be changed into Lorentzian signature as in the case of the gauged supergravity solutions.
In the ungauged case, the Wick rotation changes only the metric in the form
\begin{equation}
\fl g_{L}= \sum_{\mu=1}^2 \frac{\textrm{d}x_\mu^2}{Q_\mu}
    +\sum_{\mu=1}^2 Q_\mu \Big( \sum_{k=1}^2 \sigma^{(k-1)}_\mu d\psi_k \Big)^2
    -4\Big(\sum_{k=0}^2 \sigma^{(k)} d\psi_k+A \Big)^2 ~. \label{Lmet}
\end{equation}
The gauged minimal supergravity solutions need to correct $X_\mu$ as
\begin{eqnarray}
 X_\mu = 4 x_\mu^3+\sum_{i=1}^2 c_i x_\mu^i + b_\mu + 8 q_\mu x_\mu ~.
\end{eqnarray}
This arises from the negativity of the cosmological constant.
In both cases, the vector potential remains the form as (\ref{vec}).

%%%%%%%%%%%%%%%%%%%%%%%%%%%%%%%%%%%%%%%%%%%%%%%%%%%%%%%%%%%%%%%%%%%%%%%%%%%%%%%%%%%%%%%%%%%%%%%%%%%%
\subsection{Eleven-dimensional supergravity}
We consider the eleven-dimensional supergravity.
The action is given by 
\begin{eqnarray}
{\cal L}_{11} = *{\cal R}-\frac{1}{2}F_{(4)}\wedge* F_{(4)}
        +\frac{1}{6}F_{(4)}\wedge F_{(4)}\wedge A_{(3)}
\end{eqnarray}
where $F_{(4)}=dA_{(3)}$ is a 4-form flux of a 3-form gauge potential $A_{(3)}$.
The equations of motion are
\begin{eqnarray}
& R_{ab} = \frac{1}{12}\Big( F_{(4)}{}_{acde}F_{(4)}{}_b{}^{cde}
           -\frac{1}{12} g_{ab}F_{(4)}{}_{abcd} F_{(4)}^{abcd} \Big) ~,\label{11dEin} \\
& d* F_{(4)}-\frac{1}{2} F_{(4)} \wedge F_{(4)} = 0 ~.\label{11dCS}
\end{eqnarray}
As is the five-dimensional case, 
we examine the Euclidean solutions satisfying the Einstein equation
which are obtained by changing the sign of the right-hand side in (\ref{11dEin}).

We assume that the field strength $F_{(4)}$ takes the form
\begin{eqnarray}
 F_{(4)} = \frac{1}{2} \sum_{\mu \ne \nu} F_{\mu\nu} 
            \,e^{\mu} \wedge e^{\hat{\mu}} \wedge e^{\nu} \wedge e^{\hat{\nu}} ~, \label{F4}
\end{eqnarray}
where 
\begin{eqnarray}
 F_{\mu\nu} = 2\ell_1+\ell_2 (\partial_\mu H+\partial_\nu H) ~,
\end{eqnarray}
and $H$ is still given by (\ref{functionH}) and $\ell_1$, $\ell_2$ are constant.
Under this assumption, the field strength becomes closed, $dF_{(4)}=0$,
and the co-derivative is given by
\begin{eqnarray}
 \delta F_{(4)} 
=& -\sum_{\mu\neq\nu} \sqrt{Q_\mu}\Big(\partial_\mu F_{\mu\nu}
            +\sum_{\rho\neq\mu,\nu}\frac{F_{\mu\nu}-F_{\rho\nu}}{x_\mu-x_\rho}\Big)
            \,e^{\hat{\mu}}\wedge e^\nu\wedge e^{\hat{\nu}} \nonumber\\
 & +2 \sum_{\mu\neq\nu} F_{\mu\nu}(1+\partial_\mu H)
    \,e^\nu\wedge e^{\hat{\nu}}\wedge e^0 ~. \label{delF4}
\end{eqnarray}
Substituting the expressions (\ref{F4}) and (\ref{delF4}) into (\ref{11dCS}), 
we obtain $\ell_1=\ell_2=-2$ and
\begin{eqnarray}
& N_\mu=-x_\mu^5+\sum_{i=1}^4 a_i x_\mu^i+q_\mu ~.
\end{eqnarray}
Then we have
\begin{eqnarray}
F_{(4)} = -2\sum_{\mu\neq\nu}(1+\partial_\mu H)
          \,e^\mu\wedge e^{\hat{\mu}}\wedge e^\nu\wedge e^{\hat{\nu}} ~.
\end{eqnarray}
The Einstein equation (\ref{11dEin}) reduces to
\begin{eqnarray}
\partial^2_\mu Q_T -4 \sum_{\nu \ne \mu}K_{\mu \nu}=0 ~,
\end{eqnarray}
where 
\begin{eqnarray}
 K_{\mu \nu} \equiv -\frac{1}{4}\frac{\partial_ \mu Q_T}{x_\mu-x_\nu}
  +\frac{1}{4}\frac{\partial_\nu Q_T}{x_\mu-x_\nu} ~,~~~
 Q_T \equiv \sum_{\mu=1}^n Q_\mu ~. \label{kmunqt}
\end{eqnarray}
This equation can be solved by
\begin{eqnarray}
X_\mu = \sum_{i=1}^5 c_i x_\mu^i + b_\mu
\end{eqnarray}
with free parameters $c_i$ and $b_\mu$.

%%%%%%%%%%%%%%%%%%%%%%%%%%%%%%%%%%%%%%%%%%%%%%%%%%%%%%%%%%%%%%%%%%%%%%%%%%%%%%%%%%%%%%%%%%%%%%%%%%%%
\section{Global analysis}
%%%%%%%%%%%%%%%%%%%%%%%%%%%%%%%%%%%%%%%%%%%%%%%%%%%%%%%%%%%%%%%%%%%%%%%%%%%%%%%%%%%%%%%%%%%%%%%%%%%%

%%%%%%%%%%%%%%%%%%%%%%%%%%%%%%%%%%%%%%%%%%%%%%%%%%
\subsection{Compact manifolds in five dimensions}
%%%%%%%%%%%%%%%%%%%%%%%%%%%%%%%%%%%%%%%%%%%%%%%%%%
In this section, we discuss the global structure of
the five-dimensional minimal gauged supergravity solution obtained in section 4
and construct regular metrics on compact manifolds.

%%%%%%%%%%%%%%%%%%%%
\subsubsection{Generalization of $L^{a,b,c}$.}
%%%%%%%%%%%%%%%%%%%%
The metric is written in the form
\begin{eqnarray}
\fl \eqalign{
 g_5 =& \frac{x-y}{X}dx^2 + \frac{y-x}{Y}dy^2
         +\frac{X}{x-y}(d\psi_1+yd\psi_2)^2+\frac{Y}{y-x}(d\psi_1+xd\psi_2)^2 \\
      & +4\Big(d\psi_0+(x+y)d\psi_1+xyd\psi_2 
         + \frac{q_1-q_2}{x-y}d\psi_1+\frac{q_1y-q_2x}{x-y}d\psi_2\Big)^2 ~,
} \label{5dmet}
\end{eqnarray}
where
\begin{eqnarray}
\eqalign{
 X=-4x(x-\alpha_1)(x-\alpha_2)+b_1-8q_1 x \,, \\
 Y=-4y(y-\alpha_1)(y-\alpha_2)+b_2-8q_2 y \,.
}
\end{eqnarray}
and $\alpha_i$ $(i=1,2)$, $b_\mu$ and $q_\mu$ $(\mu=1,2)$ are free parameters.
However, not all the parameters are non-trivial. 
There is a scaling symmetry of the metric, under which we take
\begin{equation}
 \eqalign{
 x_\mu \to \lambda x_\mu ~,~~~ \psi_k \to \lambda^{-k}\psi_k ~, \\
 \alpha_i \to \lambda \alpha_i ~,~~~ b_\mu \to \lambda^3 b_\mu ~,~~~ q_i \to \lambda^2 q_\mu ~.
 } \label{scale}
\end{equation}
The metric also has a shift symmetry which is taken by
\begin{equation}
\eqalign{
 x_\mu \to x_\mu+\lambda ~,~~~ \psi_0 \to \psi_0 - \lambda^2\,\psi_2 ~,~~~
 \psi_1 \to \psi_1 -\lambda\,\psi_2 ~, \\
 \alpha_1+\alpha_2 \to \alpha_1+\alpha_2-3\lambda \,, \\
 \alpha_1\alpha_2+2q_\mu \to \alpha_1\alpha_2+2q_\mu-2(\alpha_1+\alpha_2)\lambda+3\lambda^2 ~, \\
 b_\mu \to b_\mu - 4(\alpha_1\alpha_2+2q_\mu)\lambda +4\lambda^2-4\lambda^3 ~. 
} \label{shift}
\end{equation}

In order to obtain regular metrics on compact manifolds, 
we must impose appropriate regions of the coordinates.
This corresponds to making an appropriate choice of the parameters. 
Suppose that $x_i$ and $y_i $ $(i=1,2,3)$ are real roots of the equations $X(x)=0$ and $Y(y)=0$,
and satisfy the inequalities $x_1<x_2<x_3$ and $y_1<y_2<y_3$.
If we choose the region of the coordinates as $x_1 \le x \le x_2<y_2 \le y \le y_3$, 
then the metric is positive definite, 
except for the boundaries $x=x_1$ and $x_2$ as well as $y=y_2$ and $y_3$.
From the relationship between the coefficients and solutions, we have
\begin{equation}
\eqalign{
 \alpha_1+\alpha_2 = x_1+x_2+x_3 = y_1+y_2+y_3 ~, \\
 \alpha_1\alpha_2 + 2q_1 = x_1x_2 + x_1x_3 + x_2x_3 ~, \\
 \alpha_1\alpha_2 + 2q_2 = y_1y_2 + y_1y_3 + y_2y_3 ~, \\
 b_1 = 4 x_1x_2x_3 ~,~~~ b_2 = 4 y_1y_2y_3 ~.
}
\end{equation}

Following \cite{Cvetic:2005ft,Cvetic:2009},
we can extend the metric smoothly onto the boundaries.
Since $\partial/\partial \psi_0$, 
$\partial/\partial \psi_1$ and $\partial/\partial \psi_2$ are linearly independent Killing vector fields, 
the general Killing vector field is written as
\begin{equation}
v=\sum_{k=0}^2 \omega_k \frac{\partial}{\partial \psi_k} ~,
\end{equation}
where $\omega_k$ are constants. The length of $v$ is given by
\begin{eqnarray}
\fl \eqalign{ v^2 =& \frac{X(x)}{x-y}(\omega_1+y \omega_2)^2+\frac{Y(y)}{y-x}(\omega_1+x \omega_2)^2 \\
         & +4 \Big( \omega_0+(x+y)\omega_1+xy \omega_2
           +\frac{q_1-q_2}{x-y}\omega_1+\frac{q_1y-q_2x}{x-y}\omega_2 \Big)^2 ~. }
\end{eqnarray}
Using this expression, we construct the associated normalized Killing vector fields 
$v_i$ $(i=1,2)$ and $\ell_j$ $(j=2,3)$
such that their lengths are vanishing at the corresponding boundaries $x=x_i$ and $y=y_j$.
Namely, we have
\begin{eqnarray}
& v_i=\frac{2}{X'(x_i)} \left( (q_1+x_i^2) \frac{\partial}{\partial \psi_0}
-x_i\frac{\partial}{\partial \psi_1}+\frac{\partial}{\partial \psi_2} \right) ~, \nonumber\\
& \ell_j=\frac{2}{Y'(y_j)} \left( (q_2+y_j^2) \frac{\partial}{\partial \psi_0}
-y_j\frac{\partial}{\partial \psi_1}+\frac{\partial}{\partial \psi_2} \right) ~, \label{KVs}
\end{eqnarray}
where their normalizations are taken so that the surface gravity is equal to unity,
\begin{eqnarray}
 \frac{g^{ab} (\partial_a v_i^2)( \partial_b v_i^2)}{4 v_i^2}\Big|_{x=x_i}
 = \frac{g^{ab} (\partial_a \ell_j^2) (\partial_b \ell_j^2)}{4 \ell_j^2}\Big|_{y=y_j} = 1 ~.
\end{eqnarray}
The metric extends smoothly onto the boundaries 
if the Killing vector fields $v_i$ and $\ell_j$ have period $2\pi$.

Since we have four vector fields $v_i$ and $\ell_j$,
they must satisfy a linear relation
\begin{equation}\label{reg1}
n_1 v_1+ n_2 v_2 + m_1 \ell_2 + m_2 \ell_3= 0
\end{equation}
for integral coefficients $(n_1,n_2,m_1,m_2)$, which are assumed to be coprime.
To avoid conical singularities, any three of the integers must be also coprime.
Substituting (\ref{KVs}) into (\ref{reg1}), it can be solved as
\begin{eqnarray} \label{reg2}
\fl \eqalign{
 \frac{n_1}{(x_3-x_1)[q+(x_2-y_2)(x_2-y_3)]}
  =  \frac{n_2}{(x_3-x_2)[q+(x_1-y_2)(x_1-y_3)]} \\
 =  \frac{m_1}{(y_2-y_1)[q-(x_1-y_3)(x_2-y_3)]}
  =  \frac{m_2}{(y_3-y_1)[q-(x_1-y_2)(x_2-y_2)]} ~, }
\end{eqnarray}
where
\begin{eqnarray}
 q\equiv q_1-q_2 = \frac{x_1x_2+x_1x_3+x_2x_3-y_1 y_2-y_1 y_3-y_2y_3}{2} ~.
\end{eqnarray}
Since we have degrees of freedom under the scaling symmetry (\ref{scale})
and the shift symmetry (\ref{shift}), the value of (\ref{reg2}) can be set to $1$ and we can take $b_2=0$
without loss of generality.
Then we have $y_1=0$ and (\ref{reg2}) leads to
\begin{eqnarray}
& n_1 = (x_3-x_1)[q+(x_2-y_2)(x_2-y_3)] ~, \label{reg3-1} \\
& n_2 = (x_3-x_2)[q+(x_1-y_2)(x_1-y_3)] ~, \\
& m_1 = y_2[q-(x_1-y_3)(x_2-y_3)] ~, \\
& m_2 = y_3[q-(x_1-y_2)(x_2-y_2)] ~, \label{reg3-4}
\end{eqnarray}
where
\begin{eqnarray}
 q = \frac{x_1x_2+x_1x_3+x_2x_3-y_2y_3}{2} ~. \label{reg3-5}
\end{eqnarray}
Thus the problem of constructing regular metrics on compact manifolds results in
solving four coupled algebraic equations (\ref{reg3-1})--(\ref{reg3-4}) 
for a set of coprime integers $(n_1,n_2,m_1,m_2)$,
together with the conditions for the real roots $x_i$ and $y_i$,
$x_3=y_2+y_3-x_1-x_2$, $x_1<x_2<x_3$, $0<y_2<y_3$ and $x_2<y_2$.
In particular, when we take $q = 0$, (\ref{reg3-1})--(\ref{reg3-4}) 
give rise to the condition
discussed in \cite{Cvetic:2009,Cvetic:2005ft,Martelli:2005},
\begin{equation}
n_1+n_2+m_1+m_2=0 ~,
\end{equation}
which leads to the toric Sasaki-Einstein metrics $L^{n_1, n_2, m_1}$ on $S^2 \times S^3$.
When the $q$ is non-zero, the present metric is
parameterized by independent four integers,
which we denote by $L^{n_1, n_2, m_1, m_2}$.

{\scriptsize 
\begin{center}
\begin{tabular}{|cccc|cccccc|c|c|} \hline
$n_1$ &$n_2$ &$m_1$ &$m_2$ &$x_1$ &$x_2$ &$x_3$ &$y_1$ &$y_2$ &$y_3$ &$q$  \\ \hline \hline
-4 &-3 &-1 &-2 &-1.32023 &-1.25127 &3.11486 &0 &0.167499 &0.375858 &-3.21042  \\ 
-4 &-2 &-1 &-3 &-1.27727 &-1.14007 &3.17205 &0 &0.155888 &0.59882 &-3.15254 \\
-4 &-1 &-2 &-3 &-1.25653 &-1.04966 &3.20068 &0 &0.329791 &0.564696 &-3.12434  \\ 
-4 &1 &-2 &-3 &-1.18938 &-0.78852 &3.0232 &0 &0.372652 &0.672647 &-2.6462 \\
-4 &2 &-1 &-3 &-1.11468 &-0.543466 &2.82041 &0 &0.188869 &0.9734 &-2.12735 \\
-4 &3 &-1 &-2 &-1.11385 &-0.202506 &2.46249 &0 &0.263875 &0.882254 &-1.62438 \\
-3 &-2 &-1 &-4 &-1.15876 &-1.09101 &3.24621 &0 &0.142997 &0.853444 &-3.08052 \\
-3 &-1 &-2 &-4 &-1.14654 &-1.00989 &3.26472 &0 &0.302705 &0.805588 &-3.06305 \\
-3 &1 &-2 &-4 &-1.05629 &-0.741916 &3.11358 &0 &0.3276 &0.987783 &-2.56939 \\
-3 &2 &-1 &-4 &-0.899549 &-0.461233 &3.04032 &0 &0.146529 &1.53301 &-1.97347 \\ 
-2 &-1 &-3 &-4 &-1.06256 &-0.9939 &3.29153 &0 &0.483516 &0.751555 &-3.0381 \\ 
-2 &1 &-3 &-4 &-0.969754 &-0.731764 &3.13442 &0 &0.536657 &0.896244 &-2.55231 \\ 
-1 &2 &-3 &-4 &-0.761789 &-0.47932 &2.987 &0 &0.578562 &1.16732 &-2.00871 \\ 
-1 &4 &-3 &-2 &-0.358631 &0.309102 &2.83766 &0 &0.50795 &2.28018 &-0.704807 \\ 
1 &4 &-3 &-2 &0.167966 &1.36545 &2.52543 &0 &1.90031 &2.15854 &0 \\ 
2 &1 &-3 &4 &1.58023 &2.19861 &2.46249 &0 &2.66499 &3.57634 &1.62438 \\ 
2 &1 &3 &4 &2.739 &2.94736 &3.11486 &0 &4.36613 &4.43508 &3.21042 \\ 
2 &3 &-4 &-1 &0.36689 &0.625118 &2.52543 &0 &1.15997 &2.35746 &0 \\ 
2 &3 &-4 &1 &0.557479 &2.32971 &2.83766 &0 &2.52855 &3.19629 &0.704807 \\ 
3 &1 &-2 &4 &1.84701 &2.63154 &2.82041 &0 &3.36388 &3.93509 &2.12735 \\ 
3 &1 &2 &4 &2.57323 &3.01616 &3.17205 &0 &4.31211 &4.44931 &3.15254 \\ 
3 &2 &-1 &4 &2.35055 &2.65055 &3.0232 &0 &3.81172 &4.21258 &2.6462 \\ 
3 &2 &1 &4 &2.63598 &2.87089 &3.20068 &0 &4.25033 &4.45721 &3.12434 \\ 
4 &1 &-2 &3 &1.50731 &2.8938 &3.04032 &0 &3.50156 &3.93987 &1.97347 \\ 
4 &1 &2 &3 &2.39276 &3.10321 &3.24621 &0 &4.33722 &4.40497 &3.08052 \\ 
4 &2 &-1 &3 &2.1258 &2.78598 &3.11358 &0 &3.8555 &4.16987 &2.56939 \\ 
4 &2 &1 &3 &2.45913 &2.96201 &3.26472 &0 &4.2746 &4.41126 &3.06305 \\ 
4 &3 &-2 &1 &1.81967 &2.40843 &2.987 &0 &3.46631 &3.74878 &2.00871 \\ 
4 &3 &-1 &2 &2.23817 &2.59776 &3.13442 &0 &3.86618 &4.10417 &2.55231 \\ 
4 &3 &1 &2 &2.53997 &2.80801 &3.29153 &0 &4.28543 &4.35409 &3.0381 \\ \hline
\end{tabular}\\
\parbox{10cm}{
Numerical solutions of four coupled algebraic equations (\ref{reg3-1})--(\ref{reg3-4}) 
for some sets of coprime integers $(n_1,n_2,m_1,m_2)$,
under the conditions $x_3=y_2+y_3-x_1-x_2$, $x_1<x_2<x_3$, $0<y_2<y_3$ and $x_2<y_2$
for the real roots $x_i$ and $y_i$ $(i=1,2,3)$.}
\end{center}}

%%%%%%%%%%%%%%%%%%%%
\subsection*{5.1.2. Generalization of $Y^{p,q}$}
%%%%%%%%%%%%%%%%%%%%
Making use of the five-dimensional minimal gauged supergravity solution (\ref{5dmet}),
we have discussed the global metrics on compact manifolds $M_5$
and it has been seen that they can be regarded as a generalization of $L^{a,b,c}$.
In the special case, we can also, and this time rather precisely discuss the global properties
of the metrics which can be regarded as a generalization of $Y^{p,q}$ \cite{Gauntlett:2004yd}.
Taking a certain limit of the solution (\ref{5dmet}), we obtain the metric locally given by
\begin{eqnarray}
\fl \eqalign{
 g =& (\xi-x)(d\theta^2+\sin^2 \theta d\phi^2)+\frac{dx^2}{Q(x)}+Q(x)(d\psi_1+\cos \theta d\phi)^2 \\
    & +4\left(d\psi_0+ \left(x+\frac{q}{x-\xi} \right)d\psi_1+
      \left(x-\xi+\frac{q}{x-\xi} \right)\cos \theta d\phi \right)^2 ~, } \label{5dmet2}
\end{eqnarray}
where
\begin{equation}
Q(x)=\frac{4 x^3+(1-12\xi)x^2+(8q-2\xi+12 \xi^2)x+k}{\xi-x}
\end{equation}
and $q$, $\xi$ and $k$ are free parameters.
The metric is again a Sasaki with torsion metric and satisfies the equations of motion 
of five-dimensional minimal gauged supergravity
with the Maxwell potential
\begin{equation}\label{Maxwell}
A_{(1)}=-\frac{2 \sqrt{3}q}{x-\xi}(d\psi_1+\cos \theta d\phi) ~.
\end{equation}
The torsion 3-form is given by $T=*F_{(2)}/\sqrt{3}$.

Following \cite{Gauntlett:2004yd,Gauntlett2:2006}, we study global properties of the metric (\ref{5dmet2}). 
Before starting the analysis, we perform the following coordinate transformation
\begin{eqnarray}
\psi_1=-\psi+\alpha ~,~~~ \psi_0=\xi \psi ~.
\end{eqnarray}
Then the metric is
\begin{equation} \label{5-dim.}
\fl \eqalign{
g_5 =& (\xi-x)(d\theta^2+\sin^2 \theta d\phi^2)+\frac{dx^2}{Q(x)}
       +\frac{4 \xi^2 Q(x)}{F(x)}(d\psi-\cos \theta d\phi)^2 \\
     & + F(x) \Big(d\alpha-f(x)(d\psi-\cos \theta d\phi) \Big)^2 ~, }
\end{equation} 
where 
\begin{eqnarray} 
F(x) =& Q(x)+ 4\Big( x+\frac{q}{x-\xi} \Big)^2 ~, \\
f(x) =& \frac{Q(x)+4 \left(x+\displaystyle{\frac{q}{x-\xi}} \right) 
       \left(x-\xi+\displaystyle{\frac{q}{x-\xi}} \right)}{F(x)} ~.
\end{eqnarray}
It should be noticed that when $q=0$ and $\xi=1/6$,
the metric is the local form of the Sasaki-Einstein metric $Y^{p,q}$.
In addition, we obtain the homogeneous Sasaki-Einstein metric $T^{1,1}$ if we take
the coordinate transformation $x=c/6y$ and send $c\to 0$.
We also find that when $k=-4\xi^3+\xi^2-8 q \xi$, 
the function $Q(x)$ degenerates to a polynomial of degree 2 and we have $Q=-4x^2+(8 \xi-1)x-4\xi^2+\xi-8 q$.
Then the metric is the standard $S^5$ metric when $q=0$.
Otherwise, $Q(x)$ is a rational function and henceforth we will focus on the case.

The metric $g_5$ is positive definite 
when there exist three distinct real roots $x_1$, $x_2$ and $x_3$ 
of the equation $Q(x)=0$ such that
\begin{eqnarray}\label{ineq}
x_1<x_2<x_3 ~,~~~x_2<\xi ~,
\end{eqnarray}
and the coordinate $x$ takes the range $x_1 \le x \le x_2$.
Although we will show later that the five-dimensional space $(M_5,g_5)$ 
is an $S^1$-bundle over four-dimensional space $B$ given by the metric 
\begin{eqnarray} \label{base}
\fl g_B = (\xi-x)(d\theta^2+\sin^2 \theta d\phi^2)+\frac{dx^2}{Q(x)}
       +\frac{4 \xi^2 Q(x)}{F(x)}(d\psi-\cos \theta d\phi)^2 ~,
\end{eqnarray}
we shall see first that $g_B$ can extends globally on $S^2$-bundle over $S^2$.
Fixing the coordinates ($\theta, \phi$) and
introducing a new coordinate $r= 2|(x-x_i)/Q'(x_i)|^{1/2}$,
we can evaluate the behavior near $x=x_i$ of the fiber metric as
\begin{eqnarray}\label{fiber}
dr^2+\left( \frac{\xi (x_i-\xi)Q'(x_i)}{2(x_i(x_i-\xi)+q)} \right)^2 r^2 d\psi^2 ~.
\end{eqnarray}
Hence, avoiding conical singularities at $x=x_i$ requires both of the condition
\begin{equation}\label{cond1}
\frac{\xi (x_i-\xi)Q'(x_i)}{x_i(x_i-\xi)+q}=\pm n 
\end{equation}
and the range of $\psi$ given by $ 0 \le \psi \le 4\pi/n$ with a constant $n\neq 0$.
(\ref{cond1}) is explicitly written as
\begin{eqnarray}\label{cond2}
\fl (12 \xi-n_i)x_i^2+\xi(2-24 \xi+n_i) x_i
+2\xi(4q-\xi+6 \xi^2)-q n_i=0 ~, ~~~ (i=1,2) ~,
\end{eqnarray}
where $n_i$ take $\pm n$, respectively.
Thus, two of three parameters $q$, $k$ and $\xi$ are fixed
by the regular condition (\ref{cond2}).
Since the Chern number is calculated as
\begin{equation}\label{Chern}
c_1(B)=\frac{n}{4 \pi} \int_{S^2} d(-\cos \theta d\phi)= n ~,
\end{equation}
the four-dimensional space $B$ is a trivial bundle $S^2 \times S^2$ for even integer $n$ 
and a twisted $S^2$-bundle for odd integer $n$, respectively.
For simplicity, we deal with the case $n_1=n_2=n$.
We notice that (\ref{cond2}) becomes trivial when $q=0$, $\xi=1/6$ and $n=2$,
which reproduces the Sasaki-Einstein metric $Y^{p,q}$.
In the case $\xi\neq n/12$ nor $n/16$, we obtain more general solutions of (\ref{cond2}).
\begin{eqnarray} \label{sol}
 q=\frac{(2-n)\xi(-n+4 \xi+4 n \xi)}{4(n-16 \xi)(n-12\xi)} \,, \\
 k=\frac{\xi(-n+4 \xi+8 n \xi-48 \xi^2)L(\xi)}{4(n-16 \xi)(n-12\xi)^3} \,,
\end{eqnarray}
where
\begin{eqnarray}
 L(\xi) =& 2n^2-n^3+4(n^3- n^2-6n)\xi \nonumber\\
         & +16(n^2+16 n+4)\xi^2-192( 7n+8)\xi^3+9216 \xi^4 ~.
\end{eqnarray}
Then the roots of the function $Q(x)$, $x_1$, $x_2$ and $x_3$ are given by
\begin{eqnarray}
 x_{1,2}=& \frac{2\xi+n\xi-24\xi^2\pm 
 \displaystyle{\sqrt{\frac{(n-2)n\xi(-n+5n\xi10\xi-48\xi^2)}{n-16\xi}}}}{2(n-12\xi)} ~, \label{x1 and x2} \\
 x_3 =& \frac{-n+4\xi+8n\xi-48\xi^2}{4(n-12\xi)} ~, \label{x3}
\end{eqnarray}
where the choice of the sign in (\ref{x1 and x2}) depends on the sign of $n-12\xi$.
The reality condition of $x_1$ and $x_2$ and the inequalities (\ref{ineq}) 
require the following ranges of $\xi$ for each integer $n$:
\begin{eqnarray}
 (a)& \quad n \ge 4 \,, \quad \xi_1<\xi<\frac{n}{4(n+1)} \,, \\
 (b)& \quad n=1 \,, \quad \frac{15-\sqrt{33}}{96} <\xi< \frac{1}{8} \,, \\
 (c)& \quad n\leq -1 \,, \quad \frac{n}{8}<\xi<\xi_1 \quad \textrm{or} \quad \xi_2<\xi<\xi_3 ~,
\end{eqnarray}
where the quantities $\xi_1$, $\xi_2$ and $\xi_3$ are defined by
\begin{eqnarray}
 \xi_1=&\frac{1}{96}(10+5 n - \sqrt{100-92 n+25 n^2}) ~, \\
 \xi_2=&\frac{1}{96}(10+5 n + \sqrt{100-92 n+25 n^2}) ~, \\
 \xi_3 =&\frac{1}{48}(5+ n+\sqrt{25-14 n+ n^2}) ~.
\end{eqnarray}

The regular condition for five-dimensional metric $g_5$ gives rise to further constraint,
under which we must choose the period of the fiber direction $\alpha$ in (\ref{5-dim.}) 
so as to describe a principal $S^1$-bundle over $B$.
Since the connection 1-form is given by
\begin{equation}
\mathcal{A}=f(x)(d\psi-\cos \theta d\phi)
\end{equation}
the periods $P_i~(i=1,2)$ are calculated as \cite{Gauntlett2:2006},
\begin{eqnarray}
 P_1=\frac{1}{2\pi} \int_{C_1} d \mathcal{A}=\frac{2}{n}(f(x_2)-f(x_1)) \,, \\
 P_2=\frac{1}{2\pi} \int_{C_2} d \mathcal{A}=2 f(x_2) \,,
\end{eqnarray}
where
$n$ is the Chern number given by (\ref{Chern}).
The $C_1$ and $C_2$ represent the basis for 
$H_2(B, \mathbb{Z})=\mathbb{Z} \oplus \mathbb{Z}$.
Note that the cycle $C_1$ is the $S^2$-fibre of $B$ at some fixed point $(\theta, \phi)$ on the base space,
while the $C_2$ is the sub-manifold $S^2$ of $B$ at $x=x_2$, where the length of $\partial/\partial \psi$ vanishes. 
Now we require
\begin{equation}\label{cond}
\frac{f(x_1)}{f(x_2)}=\frac{\ell}{m} ~,
\end{equation}
where $\ell, m \in \mathbb{Z}$. Then, $\kappa^{-1} d\mathcal{A}/2 \pi$  has integral periods if we set 
$\kappa=2 hf(x_2)/(m n)$
with $h=\mbox{gcd}(\ell-m, n m)$.
Thus we take the range $0 \le \alpha  \le 2 \pi \kappa$. 
A numerical calculation shows that our solution (\ref{sol}) admits 
the parameter $\xi$ satisfying the condition (\ref{cond}), and hence
the five-dimensional space $M_5$ becomes an $S^1$-bundle over $B$ 
parameterized by three integers $\ell, m$ and $n$.
It is straightforward to verify that the following four Killing vectors
\begin{eqnarray}
& v_1=\frac{\partial}{\partial \phi}+\frac{\partial}{\partial \psi},~~v_2=-\frac{\partial}{\partial \phi}+\frac{\partial}{\partial \psi} ~,\nonumber\\
& \ell_1=\frac{2}{n} \left( \frac{\partial}{\partial \psi}+ f(x_1)\frac{\partial}{\partial \alpha} \right),~~
\ell_2=\frac{2}{n} \left( \frac{\partial}{\partial \psi}+ f(x_2)\frac{\partial}{\partial \alpha} \right),~~
\end{eqnarray}
vanish  with the surface gravity 1 on the sub-manifolds given by 
$\theta=0, \theta=\pi, x=x_1$ and $x=x_2$, respectively, and they have
a linear relation 
\begin{equation}
(N_1-N_2)(v_1+v_2)+n N_2 \ell_1-n N_1 \ell_2=0
\end{equation}
with $N_1=n \ell/h \in \mathbb{Z},~N_2=n m/h \in \mathbb{Z}$ ( cf. (\ref{reg1})).\\
The volume is given by
\begin{equation}\label{vol2}
\mbox{Vol}(M_5)=\pi^3 \Bigg|\frac{32 \xi \kappa (x_2-x_1)(2 \xi-x_1-x_2)}{n}\Bigg| ~.
\end{equation}
Moreover, since $B$ is a simply-connected manifold, it follows that $M_5$ is also simply-connected.
Note also that $M_5$ is a spin manifold \cite{Gauntlett2:2006}.
Smale's theorem states that any simply-connected compact five-manifold 
which is spin and has no torsion in the second homology group is
diffeomorphic to $S^5 \sharp k(S^2 \times S^3)$ for some non-negative integer $k$. 
Thus, together with the analysis in appendix A of \cite{Gauntlett:2004yd},
we see that $M_5$ is topologically $S^2 \times S^3$.

%%%%%%%%%%%%%%%%%%%%%%%%%%%%%%%%%%%%%%%%%%%%%%%%%%%
\subsection{Non-compact manifolds in eleven dimensions}
%%%%%%%%%%%%%%%%%%%%%%%%%%%%%%%%%%%%%%%%%%%%%%%%%%
Next, we turn to discussing the global structure of the eleven-dimensional supergravity solution. 
We assume that the functions $X_\mu(x_\mu)$ take the form
\begin{eqnarray}
 X(x)     \equiv& X_1(x_1) = c(x-a)P(x) ~, \nonumber\\
 Y_k(y_k) \equiv& X_{k+1}(x_{k+1}) = c\prod_{i=1}^5(y_k-\beta_i) ~, ~~~ k=1,\cdots,4 ~,
\end{eqnarray}
where $P(x)$ is a positive definite polynomial of degree 4 and $a, c, \beta_i$ are real constants satisfying
\begin{equation}
c>0,~\beta_1<\beta_2< \cdots < \beta_5<a.
\end{equation} 
Then we choose the region of the coordinates $x, y_k$ as
\begin{equation}
\beta_1 \le y_1 \le \beta_2 \le \cdots \le y_4 \le \beta_5 < a \le x < \infty.
\end{equation}
The fact that the region of $x$ is infinite corresponds to non-compactness of manifold.
Then the metric is positive definite except for the boundaries $y_k=\beta_k, y_k=\beta_{k+1}$ and $x=a$.
From appendix B we see that the curvature is finite at the points $y_k=y_{k+1}=\beta_k$. Some calculations analogous to the five-dimensional case yield
that the following vector fields are Killing vector fields 
vanishing at the boundaries $x=a, y_k=\beta_k$ and $y_k=\beta_{k+1}$ ~(k=1,2,3,4), respectively,
\begin{eqnarray}
\fl \eqalign{
 v_0=\frac{2}{X'(a)} \left( (N_1(a)+a^5) \frac{\partial}{\partial \psi_0}+\sum_{\ell=1}^5 (-1)^k\ell a^{5-\ell} \frac{\partial}{\partial \psi_\ell} \right) ~, \\
 v_k=\frac{2}{Y_k'(\beta_k)} \left( (N_{k+1}(\beta_k)+\beta_k^5) \frac{\partial}{\partial \psi_0}+\sum_{\ell=1}^5 (-1)^\ell \beta_k^{5-\ell} \frac{\partial}{\partial \psi_\ell} \right) ~, \\
 w_k=\frac{2}{Y_k'(\beta_{k+1})} \left( (N_{k+1}(\beta_{k+1})+\beta_{k+1}^5) \frac{\partial}{\partial \psi_0}+\sum_{\ell=1}^5 (-1)^\ell \beta_{k+1}^{5-\ell} \frac{\partial}{\partial \psi_\ell} \right) ~. }
\end{eqnarray}
These Killing vector fields have a unit surface gravity. 
If we impose the condition $q_2=q_3=q_4=q_5$, 
then we have $N_2=N_3=N_4=N_5$, which implies the relation
$v_k=w_{k-1} (k=2,3,4)$. 
Hence we can use them as the new Killing coordinates $\phi_\alpha$ 
with period $2\pi$ representing the canonical coordinate of torus $T^6$,
\begin{equation}
\fl
 \frac{\partial}{\partial \phi_0}=v_0 \,, \quad \frac{\partial}{\partial \phi_1}=v_1 \,, \quad
 \frac{\partial}{\partial \phi_k}=v_k=w_{k-1} ~~(k=2,3,4) \,, \quad
 \frac{\partial}{\partial \phi_5}=\omega_4 \,, \quad
\end{equation}

\section{Summary and discussion}
Motivated by supergravity theories, we have introduced a Sasaki with torsion (ST) manifold,
which is defined as a Riemannian manifold whose metric cone is K\"ahler with torsion (KT).
In terms of almost contact metric structure,
the ST manifold is a normal almost contact metric manifold
on which the vector field $\xi$ is a Killing vector field of unit length.
The dual 1-form $\eta$ of $\xi$ is a special Killing 1-form.
Furthermore we also find special Killing forms $\eta \wedge (d^T\eta)^p$ of higher degrees.
These are all known examples of special Killing forms
at least in ordinary Sasakian manifolds except for round spheres.

In section 3, we have presented an example of the ST metric in $2n+1$ dimensions.
The metric is quasi-Sasakian and further admits $n+1$ Killing vector fields preserving the KT structure.
We also have demonstrated that there exist two kinds of hidden symmetries:
one is given by special Killing forms mentioned above 
and the other by generalized Killing-Yano (GKY) tensors
which are related to non-trivial rank-2 Killing tensors.
Although the former exists in the general ST manifold,
the existence of the GKY tensors could not always be expected.
In our case, the GKY tensors are given by the Hodge duals of
generalized closed conformal Killing-Yano (GCCKY) tensors of odd ranks:
the ST metric we presented is the first example admitting such odd-rank GCCKY tensors.
The GKY tensors lead to separation of variables in the Hamilton-Jacobi equation for geodesics.
It would be interesting to examine in this geometry whether the GKY tensors generate
separation of variables for the Klein-Gordon and Dirac equations.

Using the ST metric (\ref{met}) as an {\it ansatz},
we have constructed exact solutions in five-dimensional minimal gauged supergravity
and eleven-dimensional supergravity in section 4,
and discussed the global structures of the solutions in section 5.
The ST metrics on the five-dimensional compact manifolds provide a natural generalization of
the toric Sasaki-Einstein metrics $Y^{p,q}$ and $L^{a,b,c}$.
Indeed there exists a toric action preserving the KT structure, and the Einstein condition is replaced 
by the equations of motion of the minimal gauged supergravity.
In eleven dimensions, we have briefly analyzed the ST metrics on non-compact manifolds.
Further global analysis in eleven dimensions remains as a future problem.
We also find that a deformed $S^5$ \cite{Pilch:2000ej} and $S^7$ \cite{CPW:2002}
describing nontrivial supersymmetric solutions of supergravity theories are ST manifolds.
Therefore, it is expected that the notion of ST manifolds works well
for finding other supersymmetric solutions 
which play an important role in the AdS/CFT correspondence.

For the ST manifolds in section 2,
we have three kinds of connections with totally skew-symmetric torsion.
The first one is the Bismut connection
which preserves the KT structure of the cone.
For the five-dimensional solution of minimal gauged supergravity,
the corresponding torsion (\ref{torsion}) can be identified with the Maxwell field (\ref{Mxl}) as
\begin{eqnarray}
T=\frac{1}{\sqrt{3}}*F_{(2)} \,.
\end{eqnarray}
The second is the connection preserving the almost contact metric structure \cite{Friedrich:2002}.
The associated torsion $T_c$ is given by $T_c = T + 2\eta \wedge \omega$.
It was pointed out in the recent paper \cite{CM:2012} that this relation holds in general ST manifolds.
A supersymmetric solution in five-dimensional heterotic supergravity was discovered in \cite{Fernandez:2009},
where the three-form flux is identified with the torsion $T_c$ preserving the almost contact metric structure.
The last connection appears in the hidden symmetry of our metrics.
However, the relation between the torsion $G$ of hidden symmetry
and $T$ (or $T_c$) is not yet fully understood,
since not all of the torsion $G$ given by (\ref{GCKYtorsion}) can be expressed using the Maxwell field
or the almost contact metric structure.
It would be interesting to clarify the physical meaning of these three connections
and the relationship between them.

\section*{Acknowledgments}
We are grateful to Maciej Dunajski, Gary W.\ Gibbons, David Kubiz\v{n}\'ak and Martin Wolf
for useful comments.
T.\ H.\ also would like to thank DAMTP, University of Cambridge, for the hospitality.
The work of T.\ H.\ is supported by 
the JSPS Strategic Young Researcher Overseas Visits Program for
Accelerating Brain Circulation ``Deeping and Evolution of Mathematics and
Physics, Building of International Network Hub based on OCAMI.''
The work of Y.\ Y.\ is supported by the Grant-in Aid for Scientific Research No. 23540317
and No. 21244003 from Japan Ministry of Education.
The work of H.\ T.\ is supported by the Grant-in-Aid for the Global COE Program 
"The Next Generation of Physics, Spun from Universality and Emergence" 
from the Ministry of Education, Culture, Sports, Science and Technology (MEXT) of Japan.

\appendix
%%%%%%%%%%%%%%%%%%%%%%%%%%%%%%%%%%%%%%%%%%%%%%%%%%%%%%%%%%%%%%%%%%%%%%%%%%%%%%%%%%%%%%%%%%%%%%%%%%%%
\section{Some properties of T-contact metric manifolds}
%%%%%%%%%%%%%%%%%%%%%%%%%%%%%%%%%%%%%%%%%%%%%%%%%%%%%%%%%%%%%%%%%%%%%%%%%%%%%%%%%%%%%%%%%%%%%%%%%%%%
In this section we show some useful formulae on T-contact metric manifolds
in order to prove proposition \ref{prop4} in section 2.
Let $(M, g, T, \xi, \eta, \Phi)$ be an almost contact metric manifold 
equipped with a 3-form $T$ satisfying (\ref{torsioncond})
and define a fundamental 2-form $\omega$ by $ \omega(X,Y)=g(\Phi(X),Y)$.
Then a straightforward calculation leads us to
\begin{equation} \label{dg}
\fl \eqalign{
2 g((\nabla^T_X \Phi)(Y), Z)
 =& -d \omega(X, \Phi (Y), \Phi (Z))+d \omega(X,Y,Z)+M(X,Y,Z) \\
  & +g(N^{(1)}(Y, Z), \Phi (X))+\eta(X) N^{(2)}(Y, Z) \\
  & +d^T \eta(X, \Phi (Z)) \eta(Y)-d^T \eta(X, \Phi (Y)) \eta(Z) \,,
 }
\end{equation}
where $N^{(i)}~(i=1,2)$ are tensor fields defined in section 6 of \cite{BoyerEtAl:2008} by
\begin{eqnarray}
N^{(1)}(X,Y) =& N_{\Phi}(X,Y)+d\eta(X,Y) \xi \,, \label{N1} \\
N^{(2)}(X,Y) =& (\mathcal{L}_{\Phi (X)} \eta)(Y)-(\mathcal{L}_{\Phi (Y)} \eta)(X) \,, \label{N2}
\end{eqnarray}
and $M$ is a tensor field defined by
\begin{equation}
\eqalign{
M(X,Y,Z) =& T(X, \Phi (Y), Z)+T(X, Y, \Phi (Z)) \\
          & -T(\xi, X, \Phi (Y))\eta(Z)+T(\xi, X, \Phi (Z))\eta(Y) \,.
} 
\end{equation}
Note that
\begin{equation} \label{M}
M(\xi, X, Y)=M(X, \xi, Y)=M(X, Y, \xi)=0 \,.
\end{equation}

On a T-contact metric manifold, (\ref{dg}) simplifies.
In fact, we have 
\begin{eqnarray}
\fl \eqalign{
N^{(2)}(X,Y) 
&= d\eta(X, \Phi (Y))+d\eta(\Phi (X),Y) \\
&= d^T\eta(X, \Phi (Y))+d\eta^T(\Phi (X),Y) +T(\xi, X, \Phi (Y))+T(\xi, \Phi (X), Y) \\
&= 0 \,,
}
\end{eqnarray}
where we have used (\ref{torsioncond}), (\ref{AC4}) and (\ref{AC6}) at the last equality.
Furthermore, we obtain
\begin{equation}
\xi \hook d^T \eta=\xi \hook d \eta=0 \,.
\end{equation}
This implies that $\mathcal{L}_\xi \eta=0$ and $\mathcal{L}_\xi d \eta=0$.
It is also obtained that
\begin{eqnarray}
\mathcal{L}_\xi d^T \eta= d \xi \hook d^T \eta+\xi \hook d d^T \eta
= 2 \xi \hook d \omega = 0 \,,
\end{eqnarray}
which leads to
\begin{equation}\label{Kill}
2(\mathcal{L}_\xi g)(X,Y)= d^T \eta(X, (\mathcal{L}_\xi \Phi)(Y)) \,.
\end{equation}
Thus $\xi$ is a Killing vector field if and only if $\mathcal{L}_\xi \Phi=0$.

%%%%%%%%%%%%%%%%%%%%%%%%%%%%%%%%%%%%%%%%%%%%%%%%%%%%%%%%%%%%%%%%%%%%%%%%%%%%%%%%%%%%%%%%%%%%%%%%%%%%
\section{Some technical results}
%%%%%%%%%%%%%%%%%%%%%%%%%%%%%%%%%%%%%%%%%%%%%%%%%%%%%%%%%%%%%%%%%%%%%%%%%%%%%%%%%%%%%%%%%%%%%%%%%%%%
In this section we collect some technical results.
For the metric (\ref{met}), we compute the covariant derivatives
with respect to the Levi-Civita connection $\nabla$ in section B.1
and with respect to the connection with skew-symmetric torsion $\nabla^T$ (given in section 3.1)
in section B.2, respectively.
In section B.1, we also compute the curvature quantities with respect to $\nabla$.
The resulting curvatures have been used for solving Einstein equations
of the supergravity theories considered in section 4.

\subsection{The Levi-Civita connection}
We have chosen the orthonormal frame (\ref{ortho})
and obtained the connection 1-forms (\ref{con6})
for the metric (\ref{met}) in section 3.1.
Then, using relation $\nabla_{e_a}e^b(e_c)=-\omega^b{}_c(e_a)$,
we can compute the covariant derivatives as follows:
\begin{equation}
\eqalign{
& \nabla_{e_\mu}e_\mu
  = \sum_{\rho\neq\mu}\frac{\sqrt{Q_\rho}}{2(x_\mu-x_\rho)}\,e_\rho ~, \\
& \nabla_{e_\mu}e_\nu
  = -\frac{\sqrt{Q_\nu}}{2(x_\mu-x_\nu)}\,e_\mu ~, ~~~\mu\neq\nu \\
& \nabla_{e_\mu}e_{\hat{\mu}}
  = \sum_{\rho\neq\mu}\frac{\sqrt{Q_\rho}}{2(x_\mu-x_\rho)}\,e_{\hat{\rho}}
    -(1+\partial_\mu H)\,e_{0} ~, \\
& \nabla_{e_\mu}e_{\hat{\nu}}
  = -\frac{\sqrt{Q_\nu}}{2(x_\mu-x_\nu)}\,e_{\hat{\mu}} ~, ~~~\mu\neq\nu  \\
& \nabla_{e_{\hat{\mu}}}e_\mu
  = \partial_\mu \sqrt{Q_\mu}\,e_{\hat{\mu}}
    -\sum_{\rho\neq\mu}\frac{\sqrt{Q_\rho}}{2(x_\mu-x_\rho)}\,e_{\hat{\rho}}
    +(1+\partial_\mu H)\,e_0 ~, \\
& \nabla_{e_{\hat{\mu}}}e_\nu
  = -\frac{\sqrt{Q_\nu}}{2(x_\mu-x_\nu)}\,e_{\hat{\mu}}
    +\frac{\sqrt{Q_\mu}}{2(x_\mu-x_\nu)}\,e_{\hat{\nu}} ~, ~~~\mu\neq\nu \\
& \nabla_{e_{\hat{\mu}}}e_{\hat{\mu}}
  = -\partial_\mu \sqrt{Q_\mu}\,e_{\mu}
    +\sum_{\rho\neq\mu}\frac{\sqrt{Q_\rho}}{2(x_\mu-x_\rho)}\,e_\rho ~, \\
& \nabla_{e_{\hat{\mu}}}e_{\hat{\nu}}
  = \frac{\sqrt{Q_\nu}}{2(x_\mu-x_\nu)}\,e_{\mu}
    -\frac{\sqrt{Q_\mu}}{2(x_\mu-x_\nu)}\,e_{\nu} ~, ~~~\mu\neq\nu \\
& \nabla_{e_\mu} e_0 = (1+\partial_\mu H)\,e_{\hat{\mu}} ~, \\
& \nabla_{e_{\hat{\mu}}} e_0 = -(1+\partial_\mu H)\,e_{\mu} ~, \\
& \nabla_{e_0} e_{\mu} = (1+\partial_\mu H)\,e_{\hat{\mu}} ~, \\
& \nabla_{e_0} e_{\hat{\mu}} = -(1+\partial_\mu H)\,e_{\mu} ~, \\
& \nabla_{e_0} e_0 = 0 ~, 
} \label{CovDer1-13}
\end{equation}
where the function $H$ is given by (\ref{functionH}).

From the second structure equation
\begin{eqnarray}
 {\cal R}^a{}_b = d\omega^a{}_b + \sum_c \omega^a{}_c\wedge\omega^c{}_b \,,
\end{eqnarray}
the curvature 2-forms ${\cal R}^a{}_b$ are obtained as follows:
\begin{equation}
\fl \eqalign{
{\cal R}^\mu{}_\nu 
=& K_{\mu \nu}
    \,e^\mu \wedge e^\nu 
   +\Big(K_{\mu \nu}-(1+\partial_\mu H)(1+\partial_\nu H) \Big)
    \,e^{\hat{\mu}}\wedge e^{\hat{\nu}} \\
 & -\frac{\partial_\mu H-\partial_\nu H}{2(x_\mu-x_\nu)} \sqrt{Q}_\nu
    \,e^{\hat{\mu}}\wedge e^0
   +\frac{\partial_\mu H-\partial_\nu H}{2(x_\mu-x_\nu)} \sqrt{Q}_\mu
    \,e^{\hat{\nu}}\wedge e^0,~~~(\mu \ne \nu) \\
 {\cal R}^\mu{}_{\hat{\mu}}
=& -\frac{1}{2}\Big(\partial_\mu^2Q_T+6(1+\partial_\mu H)^2\Big)
    \,e^\mu \wedge e^{\hat{\mu}} \\
 & +2 \sum_{\nu \ne \mu}\Big(K_{\mu \nu}-(1+\partial_\mu H)(1+\partial_\nu H)\Big)
    \,e^\nu \wedge e^{\hat{\nu}} \\
 & -\sqrt{Q}_\mu \partial_\mu^2 H
    \,e^\mu \wedge e^0
   -\sum_{\nu \ne \mu} \frac{\partial_\mu H-\partial_\nu H}{x_\mu-x_\nu}\sqrt{Q}_\nu
    \,e^\nu \wedge e^0 \\
 {\cal R}^\mu{}_{\hat{\nu}} 
=& K_{\mu \nu}
    \,e^\mu \wedge e^{\hat{\nu}} 
   +\Big(K_{\mu \nu}-(1+\partial_\mu H)(1+\partial_\nu H) \Big)
    \,e^\nu \wedge e^{\hat{\mu}} \\
 & -\frac{\partial_\mu H-\partial_\nu H}{2(x_\mu-x_\nu)} \sqrt{Q}_\nu 
    \,e^\mu \wedge e^0
   -\frac{\partial_\mu H-\partial_\nu H}{2(x_\mu-x_\nu)} \sqrt{Q}_\mu 
    \,e^\nu \wedge e^0,~~~(\mu \ne \nu) \\
 {\cal R}^{\hat{\mu}}{}_{\hat{\nu}} 
=& K_{\mu \nu}
    \,e^{\hat{\mu}} \wedge e^{\hat{\nu}} 
   +\Big(K_{\mu \nu}-(1+\partial_\mu H)(1+\partial_\nu H) \Big)
    \,e^{\mu}\wedge e^\nu \\
 & -\frac{\partial_\mu H-\partial_\nu H}{2(x_\mu-x_\nu)} \sqrt{Q}_\nu 
    \,e^{\hat{\mu}} \wedge e^0
   +\frac{\partial_\mu H-\partial_\nu H}{2(x_\mu-x_\nu)} \sqrt{Q}_\mu 
    \,e^{\hat{\nu}} \wedge e^0 ~,~~~ (\mu \ne \nu) \\ 
  {\cal R}^\mu{}_0
=& -\sqrt{Q}_\mu \partial_\mu^2 H 
    \,e^\mu \wedge e^{\hat{\mu}}
   -\sum_{\nu \ne \mu} \frac{\partial_\mu H-\partial_\nu H}{2(x_\mu-x_\nu)} \sqrt{Q}_\nu 
    \,e^\nu \wedge e^{\hat{\mu}} \\
 & - \sum_{\nu \ne \mu} \frac{\partial_\mu H-\partial_\nu H}{2(x_\mu-x_\nu)} \sqrt{Q}_\nu 
    \,e^\mu \wedge e^{\hat{\nu}} 
   -\sum_{\nu \ne \mu} \frac{\partial_\mu H-\partial_\nu H}{x_\mu-x_\nu} \sqrt{Q}_\mu
    \,e^\nu \wedge e^{\hat{\nu}} \\
 &+(1+\partial_\mu H)^2 
    \,e^\mu \wedge e^0 \\
 {\cal R}^{\hat{\mu}}{}_0
=& -\sum_{\nu \ne \mu} \frac{\partial_\mu H- \partial_\nu H}{2(x_\mu-x_\nu)} \sqrt{Q}_\nu 
    \,e^\mu \wedge e^\nu
   -\sum_{\nu \ne \mu} \frac{\partial_\mu H- \partial_\nu H}{2(x_\mu-x_\nu)} \sqrt{Q}_\mu 
    \,e^{\hat{\mu}} \wedge e^{\hat{\nu}} \\
 & +(1+\partial_\mu H)^2
    \,e^{\hat{\mu}} \wedge e^0, 
    }
\end{equation}
where $K_{\mu\nu}$ and $Q_T$ are given by (\ref{kmunqt}).
The Ricci curvature is defined by
\begin{eqnarray}
 Ric(e_a,e_b) = \sum_c {\cal R}^c{}_a(e_c,e_b) \,.
\end{eqnarray}
Thus nonzero components of the Ricci curvature are
\begin{eqnarray}
\fl \eqalign{
 Ric(e_\mu,e_\mu) 
=& Ric(e_{\hat{\mu}},e_{\hat{\mu}})
=-\frac{1}{2}\partial^2_\mu Q_T+2 \sum_{\nu \ne \mu}K_{\mu \nu}-2(1+\partial_\mu H)^2 ~, \\
 Ric(e_0,e_0)
=& 2\sum_{\mu=1}^n(1+\partial_\mu H)^2 ~, \\
 Ric(e_0,e_{\hat{\mu}})
=& -\sqrt{Q_\mu} \Bigg(\partial_\mu^2 H 
   +\sum_{\nu\ne \mu}\frac{\partial_\mu H-\partial_\nu H}{x_\mu-x_\nu}\Bigg) ~.
   }
\end{eqnarray}
The scalar curvature is defined by
\begin{eqnarray}
 scal = \sum_a Ric(e_a,e_a) ~.
\end{eqnarray}
Thus we obtain
\begin{eqnarray}
 scal 
= -\sum_{\mu=1}^n \partial_\mu^2 Q_T + 4 \sum_{\mu \ne \nu} K_{\mu \nu} 
 -2\sum_{\mu=1}^n (1+\partial_\mu H)^2 ~.
\end{eqnarray}

\subsection{The connection with totally skew-symmetric torsion}
We next compute the covariant derivatives with respect to the torsion $\nabla^T$.
with respect to the orthonormal frame (\ref{ortho}) of the metric (\ref{met}).
Since we have obtained the covariant derivatives 
with respect to the Levi-Civita connection $\nabla$, (\ref{CovDer1-13}),
hence we can compute from (\ref{eq00}) the covariant derivatives
with respect to the torsion connection $\nabla^T$ as
\begin{eqnarray}
 \nabla^T_{e_a}e_b = \nabla_{e_a}e_b +\frac{1}{2} T(e_a,e_b) ~.
\end{eqnarray}
Thus we obtain
\begin{equation}
\eqalign{
& \nabla^T_{e_\mu}e_\mu
  = \sum_{\rho\neq\mu}\frac{\sqrt{Q_\rho}}{2(x_\mu-x_\rho)}\,e_\rho ~, \\
& \nabla^T_{e_\mu}e_\nu
  = -\frac{\sqrt{Q_\nu}}{2(x_\mu-x_\nu)}\,e_\mu ~, \\
& \nabla^T_{e_\mu}e_{\hat{\mu}}
  = \sum_{\rho\neq\mu}\frac{\sqrt{Q_\rho}}{2(x_\mu-x_\rho)}\,e_{\hat{\rho}}-e_0 ~, \\
& \nabla^T_{e_\mu}e_{\hat{\nu}}
  = -\frac{\sqrt{Q_\nu}}{2(x_\mu-x_\nu)}\,e_{\hat{\mu}} ~,  \\
& \nabla^T_{e_{\hat{\mu}}}\,e_\mu
  = \partial_\mu \sqrt{Q_\mu}\,e_{\hat{\mu}}
    -\sum_{\rho\neq\mu}\frac{\sqrt{Q_\rho}}{2(x_\mu-x_\rho)}\,e_{\hat{\rho}}
    +e_0 ~, \\
& \nabla^T_{e_{\hat{\mu}}}e_\nu
  = -\frac{\sqrt{Q_\nu}}{2(x_\mu-x_\nu)}\,e_{\hat{\mu}}
    +\frac{\sqrt{Q_\mu}}{2(x_\mu-x_\nu)}\,e_{\hat{\nu}} ~, \\
& \nabla^T_{e_{\hat{\mu}}}\,e_{\hat{\mu}}
  = -\partial_\mu \sqrt{Q_\mu}\,e_{\mu}
    +\sum_{\rho\neq\mu}\frac{\sqrt{Q_\rho}}{2(x_\mu-x_\rho)}\,e_\rho ~, \\
& \nabla^T_{e_{\hat{\mu}}}e_{\hat{\nu}}
  = \frac{\sqrt{Q_\nu}}{2(x_\mu-x_\nu)}\,e_{\mu}
    -\frac{\sqrt{Q_\mu}}{2(x_\mu-x_\nu)}\,e_{\nu} ~, \\
& \nabla^T_{e_\mu} e_0 = e_{\hat{\mu}} ~, \\
& \nabla^T_{e_{\hat{\mu}}} e_0 = -e_{\mu} ~, \\
& \nabla^T_{e_0} e_{\mu} = (1+2 \partial_\mu H)\,e_{\hat{\mu}} ~, \\
& \nabla^T_{e_0} e_{\hat{\mu}} = -(1+2 \partial_\mu H)\,e_{\mu} ~, \\
& \nabla^T_{e_0} e_0 = 0 ~,  
} \label{CovDer2-13}
\end{equation}
where the function $H$ is again given by (\ref{functionH}).

%%%%%%%%%%%%%%%%%%%%%%%%%%%%%%%%%%%%%%%%%%%%%%%%%%%%%%%%%%%%%%%%%%%%%%%%%%%%%%%%%%%%%%%%%%%%%%%%%%%%
\section{Calabi-Yau with torsion metric on a cone}
%%%%%%%%%%%%%%%%%%%%%%%%%%%%%%%%%%%%%%%%%%%%%%%%%%%%%%%%%%%%%%%%%%%%%%%%%%%%%%%%%%%%%%%%%%%%%%%%%%%%
We begin with the metric (\ref{met_cone}) and choose the same orthonormal frame as (\ref{ortho_cone}),
then the connection 1-forms are calculated as (\ref{coneomega}). 
For the Hermitian connection $\bar{\nabla}^B$ 
with respect to the Bismut torsion (\ref{B_cone}),
the connection 1-form with torsion $\bar{\omega}^B{}^\alpha{}_\beta$ are calculated as
\begin{eqnarray}
 \bar{\omega}^B{}^\alpha{}_\beta 
= \bar{\omega}^\alpha{}_\beta 
  -\frac{1}{2}\sum_\gamma B^\alpha{}_{\beta\gamma}\,\bar{e}^\gamma ~.
\end{eqnarray}
That is, we have
\begin{eqnarray}
\eqalign{
 \bar{\omega}^B{}^r{}_a 
=& -\frac{\bar{e}^a}{r} ~, \\
 \bar{\omega}^B{}^\mu{}_\nu 
=& -\frac{\sqrt{Q_\nu}}{2(x_\mu-x_\nu)}\,\frac{\bar{e}^\mu}{r}
   -\frac{\sqrt{Q_\mu}}{2(x_\mu-x_\nu)}\,\frac{\bar{e}^\nu}{r} ~, ~~~ \mu\neq\nu \\
 \bar{\omega}^B{}^\mu{}_{\hat{\mu}} 
=& -\partial_\mu\sqrt{Q_\mu}\,\frac{\bar{e}^{\hat{\mu}}}{r}
   +\sum_{\nu\neq\mu}\frac{\sqrt{Q_\nu}}{2(x_\mu-x_\nu)}\,\frac{\bar{e}^{\hat{\nu}}}{r}
   -(1+2\partial_\mu H)\,\frac{\bar{e}^0}{r} ~, \\
 \bar{\omega}^B{}^\mu{}_{\hat{\nu}} 
=& \frac{\sqrt{Q_\nu}}{2(x_\mu-x_\nu)}\,\frac{\bar{e}^{\hat{\mu}}}{r}
   -\frac{\sqrt{Q_\mu}}{2(x_\mu-x_\nu)}\,\frac{\bar{e}^{\hat{\nu}}}{r} ~, ~~~ \mu\neq\nu \\
 \bar{\omega}^B{}^{\hat{\mu}}{}_{\hat{\nu}}
=& -\frac{\sqrt{Q_\nu}}{2(x_\mu-x_\nu)}\,\frac{\bar{e}^\mu}{r}
   -\frac{\sqrt{Q_\mu}}{2(x_\mu-x_\nu)}\,\frac{\bar{e}^\nu}{r} ~, ~~~ \mu\neq\nu \\
 \bar{\omega}^B{}^\mu{}_0 
=& -\frac{\bar{e}^{\hat{\mu}}}{r} ~, \\
 \bar{\omega}^B{}^{\hat{\mu}}{}_0 
=& \frac{\bar{e}^\mu}{r} ~.
}
\end{eqnarray}
Note that if we restrict the connection 1-forms $\bar{\omega}^B{}^\alpha{}_\beta$ on the hyperplane of $r=1$,
then we obtain the connection 1-form $\omega^T{}^a{}_b=\bar{\omega}^B{}^a{}_b\big|_{r=1}$
with respect to the original metric $g^{(2n+1)}$ and the torsion $T$.
Since the curvature 2-form $ \bar{{\cal R}}^B{}^\alpha{}_\beta$
and the Ricci form $\rho^B(X,Y)$ are given \cite{Ivanov:2001} as
\begin{eqnarray}
& \bar{{\cal R}}^B{}^\alpha{}_\beta (X,Y)= g(\bar{R}^B(X,Y)\bar{e}_\alpha,\bar{e}_\beta) ~, \\
& \rho^B(X,Y)
   = \frac{1}{2}\sum_\alpha \bar{{\cal R}}^B(X,Y,\bar{e}_\alpha,J(\bar{e}_{\alpha})) ~,
\end{eqnarray}
where $\bar{R}^B(X,Y)$ is the curvature defined by (\ref{curvop})
with respect to $\bar{\nabla}^B$, we have the curvature 2-form as 
\begin{equation}
\fl \eqalign{
 \bar{{\cal R}}^B{}^r{}_0
=& -2\sum_{\mu=1}^n \partial_\mu H 
   \,e^\mu \wedge e^{\hat{\mu}} ~, \\
 \bar{{\cal R}}^B{}^\mu{}_{\hat{\mu}}
=& -\frac{1}{2}\Big( \partial^2_\mu Q_T+4\Big)
   \,e^\mu \wedge e^{\hat{\mu}}
   +\frac{1}{2}\sum_{\nu \ne \mu} \left(-\frac{\partial_\mu Q_T}{x_\mu-x_\nu} 
   +\frac{\partial_\nu Q_T}{x_\mu-x_\nu} \right)
   \,e^\nu \wedge e^{\hat{\nu}} \\
 & -2\sum_{\nu=1}^n \sqrt{Q}_\nu \partial_\mu\partial_\nu H
   \,e^\nu \wedge e^0 -2\sum_{\nu=1}^n(1+2 \partial_\mu H)(1+\partial_\nu H)
   \,e^\nu \wedge e^{\hat{\nu}} ~,
   }
\end{equation}
and the non-zero components of the Ricci form as
\begin{equation}
\eqalign{ \rho^B(e_\mu, e_{\hat{\mu}})
=& -\frac{1}{2} \partial^2_\mu Q_T
     +\frac{1}{2} \sum_{\nu \ne \mu} \Big( -\frac{\partial_\mu Q_T}{x_\mu-x_\nu}
     +\frac{\partial_\nu Q_T}{x_\mu-x_\nu}\Big) \\
   & -2(n+1)(1+\partial_\mu H) \,. }
\end{equation}
Thus we find that $\rho^B(X,Y)=0$ for all vector fields $X,Y$, when provided that
the functions $X_\mu$ and $N_\mu$ take the form
\begin{eqnarray}
X_\mu(x_\mu) =& -4x_\mu^{n+1}+\sum_{j=1}^nc_jx_\mu^j + b_\mu - 4(n+1)q_\mu x_\mu ~, \label{CCC} \\
N_\mu(x_\mu) =& \sum_{i=1}^{n-1} a_i x_\mu^i + q_\mu ~,
\end{eqnarray}
where $a_i$, $b_j$, $m_\mu$ and $q_\mu$ are constant parameters.
This gives a Calabi-Yau with torsion metric on a cone.
The function (\ref{CCC}) in five dimensions is different from (\ref{eq4-10a}).

\end{document}